\documentstyle[12pt]{article}
\begin{document}
\textheight 200mm
\textwidth 160mm
\leftmargin 10mm
 
\begin {center}
\bf
{\bf  Effective $\bf kp$-Hamiltonian and correct boundary conditions
for envelope functions
 in $A_3B_5$-heterostructures. Method of invariants.}
\end {center}

\vspace{0.5cm}
\begin{center}
 {\bf G.F. Glinskii, K.O. Kravchenko\cite{*}}
\end{center}

\begin{center}
\it
Saint-Petersburg Electrotechnical University, 197376 St.Petersburg, Russia 
\end {center}

\begin {abstract}
Method of invariants is used to obtain effective ${\bf kp}$-Hamiltonian 
with position-dependent band parameters and correct boundary 
conditions for electron and hole envelope functions in 
$A_3B_5$-heterostructures with arbitrary interface orientation. It is shown 
that the presence of heterointerface yields  additional 
quadratic-in-${\bf k}$ and linear-in-${\bf k}$ 
terms in the kinetic energy operator of 
electron and hole due to non-commutativity of position-dependent 
band parameters and momentum operator. In particular, the $\Gamma_6$ 
conduction band Hamiltonian is determined by two position-dependent 
band parameters: $\alpha_1 = m_0 / m_e$ and additional parameter $\alpha_2$. 
Similarly for 
description of the $\Gamma_8$ valence band ${\bf kp}$-Hamiltonian the 
three Luttinger parameters $\gamma_1$, $\gamma_2$ and $\gamma_3$ are 
no longer sufficient, and introducing of two additional parameters  
$\gamma_4$ and $\gamma_5$, that determine quadratic-in-${\bf k}$ terms, 
and additional parameter $\beta_1$, determining a linear-in-${\bf k}$ term, 
is necessary. The additional terms in the $\Gamma_6$ and $\Gamma_8$ 
${\bf kp}$-Hamiltonians, which are proportional to $\alpha_2$ and 
$\gamma_4$, $\gamma_5$, appear due to interaction at the interface 
of electron spin and hole effective spin with their orbital motion, 
which is described by envelope functions (interface effective spin-orbit 
interaction). The parameters $\alpha_2$ and $\gamma_5$  arise 
due to relativistic effects and appear in the second order of 
${\bf kp}$-perturbation theory only when spin-orbit splitting of intermediate 
states is taken into account. The parameter $\gamma_4$  is non-relativistic 
and can be approximately related to the Luttinger parameters as 
$\gamma_4 = (1/3)(-1-\gamma_1 +2\gamma_2+3\gamma_3)$. The additional 
linear-in-${\bf k}$ term, which is determined by the parameter $\beta_1$, 
is also non-relativistic and arises due to variation of the Bloch functions 
across the interface. In the framework of this approach the 
$\Gamma_8 {\oplus}\Gamma_7$
 valence band ${\bf kp}$-Hamiltonian is derived. Presence of heterointerface 
gives rise to short range interface $\delta$-corrections, which are also 
determined by the method of invariants. The new form of the Hamiltonian 
allows us to obtain the correct boundary conditions for envelope functions 
of electron and hole in $A_3B_5$-heterostructures with arbitrary 
orientated heterointerface.
\vspace{0.5cm}
\end {abstract}

\begin {center}
\bf
I. INTRODUCTION
\end {center}

In spite of the fact that the effective-mass theory for semiconductor
 heterostructures was being widely discussed in the literature 
\cite{1}-\cite{8}, by now there is no unified opinion about the proper 
form of ${\bf kp}$-Hamiltonian and corresponding boundary conditions 
for electron and hole envelope functions \cite{9}-\cite{12}.  In the present 
work it is shown that similarly to the case of bulk material the effective 
mass Hamiltonian in a system with heterointerface can be determined in 
the framework of ${\bf kp}$-perturbation theory by means of the finite number 
of band parameters. However, due to absence of a translational symmetry 
these parameters become position-dependent and, generally, their number 
exceeds the number of similar parameters in the bulk material \cite{13}. 
Besides that the presence of interface gives rise to additional short 
range interface $\delta$-corrections, which are, in particular, responsible 
for mixing of light and heavy hole states in the $\Gamma_8$ valence 
band \cite{14}. 
The approach used in this paper is close to the one used by Karavaev and 
Tikhodeev \cite{3} and by Foreman \cite{8},  \cite{10}.

Using the method of invariants we have shown that in contrast to the 
bulk material in a heterointerface system the kinetic energy operator 
of electron in the $\Gamma_6$ band up to quadratic-in-${\bf k}$ terms 
is determined by two position-dependent parameters $\alpha_1$ and $\alpha_2$. 
The first one is the inverse effective mass of electron in the $\Gamma_6$ 
band:
$\alpha_1 = m_o /m_e$. The second $\alpha_2$ is an additional parameter 
that determines a correction to the effective Hamiltonian which is 
responsible for electron spin-orbit interaction at the interface, 
calculated by means of envelope functions. It arises in the second order of 
$\bf kp$-perturbation theory when spin-orbit splitting of intermediate 
states is taken into account. Neglecting ${\bf kp}$-interaction of the 
$\Gamma_6(\Gamma_1^c)$ conduction band with all other states, except the 
nearest $\Gamma_8 {\oplus}\Gamma_ 7(\Gamma_{15}^v)$ valence band, 
this relativistic parameter 
is proportional to the $\Gamma_{15}^v$ valence band spin-orbit splitting 
$\Delta_{so}$:
$\alpha_2 = [\Delta_{so}/(3E_g+2\Delta_{so})](1-m_0/m_e)$, where 
$E_g = E_c^{\Gamma_6} - E_v^{\Gamma_8}$, 
$\Delta_{so} = E_v^{\Gamma_8} - E_v^{\Gamma_7}$.

Similarly it is shown that the hole kinetic energy operator 
in the $\Gamma_8$ 
band up to the second power of ${\bf k}$ accuracy (excluding linear 
relativistic terms) is determined by six position-dependent parameters. 
Three of them $\gamma_1$, $\gamma_2$ and $\gamma_3$ coincide with the 
usual Luttinger parameters for the bulk material, the three additional 
ones $\gamma_4$, $\gamma_5$  and $\beta_1$ arise due to the 
presence of heterointeface. The additional quadratic-in-${\bf k}$ terms in 
the $\Gamma_8$ band ${\bf kp}$-Hamiltonian, which are proportional to 
$\gamma_4$ and $\gamma_5$, appear due to interaction of hole effective
spin ($J = 3/2$) with it's orbital momentum at the interface, calculated 
by means of smooth envelope functions. The relativistic parameter 
$\gamma_5$ is 
analogous to the parameter $\alpha_2$ of the $\Gamma_6$ band and if we take 
into account interaction of the $\Gamma_8(\Gamma_{15}^v)$ valence band only 
with the nearest $\Gamma_8 {\oplus}\Gamma_7(\Gamma_{15}^c)$ 
conduction band it 
is proportional to the spin-orbit splitting  $\Delta_{so}^\prime$ of 
the $\Gamma_{15}^c$ conduction band: 
$\gamma_5 = (1/18)(1+\gamma_1-2\gamma_2)
\left[ \Delta_{so}^\prime/(E_g^\prime+\Delta_{so}^\prime) \right]$, where 
$E_g^\prime = E_c^{\Gamma_7}-E_v^{\Gamma_8}$,
$\Delta_{so}^\prime = E_c^{\Gamma_8} - E_c^{\Gamma_7}$. 
The non-relativistic parameter 
$\gamma_4$ can be approximately related to the Luttinger parameters as:
$\gamma_4 = (1/3)(-1-\gamma_1+2\gamma_2+3\gamma_3)$. The parameter $\beta_1$
 determines an additional linear-in-${\bf k}$ non-relativistic term and 
arises due to variation of the Bloch functions across interface.

The additional quadratic-in-${\bf k}$ terms in the $\Gamma_6$ and 
$\Gamma_8$ {\bf kp}-Hamiltonians, that are proportional to the parameters
$\alpha_2$ and $\gamma_4$, $\gamma_5$, which are absent in the bulk 
material, take place only in the non-homogenous systems with disturbed 
translational symmetry. These terms describe interaction of electron 
spin and hole effective spin with their orbital 
momentum, calculated by means of
smooth envelope functions, in the narrow interface region. This 
interface effective spin-orbit interaction takes place in the $\Gamma_{15}$ 
valence band, where the hole should be attributed effective spin 
$J=1$. It explains non-relativistic background of the parameter 
$\gamma_4$ in the $\Gamma_8$ ${\bf kp}$-Hamiltonian.

This approach was used to obtain the hole
$\Gamma_8 {\oplus} \Gamma_7 (\Gamma_{15}^v) $
two-band ${\bf kp}$-Hamiltonian in heterosystem. It was shown that 
if we consider the $\Gamma_8 {\oplus} \Gamma_7$ valence band states as 
formed exclusively by the $\Gamma_{15}^v$ state, i.e. neglecting spin-orbit 
mixing of these states with other states, quadratic-in-${\bf k}$ terms of the 
$6\times6$ hole ${\bf kp}$-Hamiltonian are determined by five 
position-dependent 
parameters $\gamma_1$, $\gamma_2$, $\gamma_3$ and $\gamma_4$.
Besides that the $\Gamma_8 {\oplus} \Gamma_7$ valence band 
${\bf kp}$-Hamiltonian includes a non-relativistic linear-in-${\bf k}$ term 
proportional to the parameter $\beta_1$ of the $\Gamma_8$ band.

Taking into account short range part of the interface potential gives 
rise to additional interface short range $\delta$-corrections in the 
effective mass Hamiltonian, which also include ${\bf k}$-dependent terms. 
These corrections modify boundary conditions for envelope functions 
and their derivatives and lead to additional spin-dependent effects 
of electron and hole interface scattering. Using the method 
of invariants we can determine the form of these corrections for the 
case of arbitrary orientated heterointerface.

The effective ${\bf kp}$-Hamiltonians obtained in this paper allow 
us to derive the correct boundary conditions for electron and hole 
envelope functions in $A_3B_5$-heterostructures with arbitrary 
orientated heterointerface. As an example we 
obtained boundary conditions for heterostructure with 
(001) interface.

\begin {center}
\bf
II. EFFECTIVE MASS APPROXIMATION IN HETEROINTERFACE 
SYSTEM. POSITION-DEPENDENT PSEUDO-COORDINATE BASIS
\end {center}

In contrast to the bulk material, a system with heterointerface does 
not have a translational symmetry and therefore electron wave vector 
${\bf k}$ is not a good quantum number any more. 
The electron and hole scattering 
processes at the heterointerface, changing the wave vector from ${\bf k}$ to 
${\bf k^\prime}$, lead to the fact that $\bf kp$-Hamiltonian of the 
heterosystem becomes dependent on two variables $\bf k$ and $\bf k^\prime$.

Let's represent the Hamiltonian of a system with heterointerface in the 
following form:

$$
H=\frac{{\bf p}^2}{2m_0}+U_1({\bf x})+f({\bf x})\cdot 
\Delta U({\bf x}).\eqno (1)
$$ 

\par \noindent
Here $U_{1,2}({\bf x})$ are the crystal potentials on the left 
and right sides of the interface respectively, which have the same 
period and symmetry; $\Delta U({\bf x}) = U_2({\bf x}) - U_1({\bf x})$;
$f({\bf x}) \equiv f({\bf nx})$  is a step-like function, which modulates  
periodic potential in the direction perpendicular to the interface; 
$\bf n$ is a unit vector which is normal to the interface. Usually to solve 
Schr{\"o}dinger equation with the Hamiltonian (1), the 
Kohn-Luttinger basis  
$\left| \alpha , n, {\bf k} \right\rangle = 
e^{i{\bf kx}} \left|\alpha , n \right\rangle$ is used, where 
$\left| \alpha , n \right\rangle$ is a Bloch state at ${\bf k} = 0$, 
satisfying the Schr{\"o}dinger equation for the bulk crystal with 
periodic potential $U_1({\bf x})$

$$
\left[\frac{{\bf p}^2}{2m_0}+U_1({\bf x}) \right] 
\left| \alpha , n \right\rangle 
= E^{(\alpha)}\left|\alpha ,n \right\rangle ,
$$

\par \noindent
where $\alpha$ labels bands (including irreducible representations) 
at the $\Gamma$ point of Brillouin zone, and $n$ refers to degenerate 
states (partners). In the Kohn-Luttinger basis the Hamiltonian (1) 
has the following form:

$$
\begin {array}{c}
\left\langle \alpha ,n, {\bf k}\right| H \left| 
\alpha^\prime, n^\prime, {\bf k}^\prime \right\rangle = 
\left( E^{(\alpha)} + \frac{\hbar^2k^2}{2m_0} \right) 
\delta_{\alpha \alpha^\prime}
\delta_{n n^\prime} \delta_{{\bf k} {\bf k}^\prime}+\\
f({\bf k}- {\bf k}^\prime) \Delta U^{\alpha \alpha^\prime} 
\delta_{n n^\prime}+
\frac{\hbar}{m_0} {\bf k}{\bf p}^{\alpha \alpha^\prime}_{n n^\prime} 
\delta_{{\bf k}{\bf k}^\prime} + V^{\alpha \alpha^\prime}
_{n n^\prime}({\bf k}-{\bf k}^\prime),
\end{array}
\eqno (2)
$$

\par \noindent 
where $f({\bf k}-{\bf k}^\prime) = 
f( k_{\bot}- k^\prime_{\bot}) 
\delta_{{\bf k}_{\|}{\bf k}_{\|}^\prime}$ is 
a Fourier transformant of $f({\bf x})$, $k_{\bot} = {\bf nk}$, 
${\bf k}_{\|} = {\bf k} - {\bf n}k_{\bot}$; 
${\bf p}^{\alpha \alpha^\prime}_{n n^\prime} =
\left\langle \alpha , n \right| {\bf p} \left| \alpha^\prime 
, n^\prime \right\rangle $ ; 
$\Delta U^{\alpha \alpha^\prime} \delta_{n n^\prime} =
\Delta U^{\alpha \alpha^\prime}_{n n^\prime}(0)$ are matrix elements 
which determine smooth part of the interface scattering potential 
and couple states of the same symmetry;
$V^{\alpha \alpha^\prime}_{n n^\prime}({\bf k}-{\bf k}^\prime)
= V^{\alpha \alpha^\prime}_{n n^\prime} 
\delta_{{\bf k}_{\|} {\bf k}_{\|}^\prime}$  are interface short range
corrections;
$ V^{\alpha \alpha^\prime}_{n n^\prime} =
\sum\limits_{G_{\bot} \ne 0}
f(G_{\bot}) \Delta U ^{\alpha \alpha^\prime}_{n n^\prime} (G_{\bot},
 {\bf G}_{\|}=0)$ ; 
$\Delta U ^{\alpha \alpha^\prime}_{n n^\prime} ({\bf G}) =
(1/\Omega) \int\limits_\Omega u^{\alpha^*}_{n 0} ({\bf x}) \Delta U ({\bf x}) 
u^{\alpha^\prime}_{n^\prime 0} ({\bf x}) e^{i{\bf Gx}} d{\bf x} $ ; $\bf G$ 
 is the reciprocal lattice vector, $G_{\bot} = {\bf nG}$, 
${\bf G}_{\|} = {\bf G} - {\bf n}G_{\bot}$; 
$(1/ \sqrt V) u^{\alpha }_{n 0} ({\bf x}) = 
\left\langle {\bf x} | \alpha , n \right\rangle$   
are Bloch wave functions at ${\bf k} = 0$, 
$V$  is the crystal volume, $\Omega$  is the primitive cell volume. 

To apply the effective mass approximation to heterointerface systems it is 
convenient to use pseudo-Wannier basis 
$\left| \alpha, n, {\bf R}_m \right\rangle$, which is related to the 
Kohn-Luttinger 
basis $\left| \alpha, n, {\bf k} \right\rangle$ by a unitary transformation 

$$
\left| \alpha, n, {\bf R}_m \right\rangle = \sqrt {\frac {\Omega}{V}}
 \sum\limits_{\bf k}^{BZ} \left| \alpha, n, 
{\bf k} \right\rangle e^{-i{\bf k}{\bf R}_m}
, \eqno (3)
$$

\par \noindent 
where ${\bf R}_m$ are discrete Bravais lattice vectors and 
the summation is performed over Brillouin zone. This basis 
satisfies the conditions of orthonormality and completeness

$$
\left\langle \alpha, n, {\bf R}_m | \alpha^\prime, n^\prime, 
{\bf R}_{m^\prime }
 \right\rangle = \delta_{\alpha  \alpha^\prime }\delta_{n n^\prime }
\delta_{{\bf R}_m {\bf R}_{m^\prime }},
$$

$$
\sum\limits_{\alpha, n, {\bf R}_m } \left| \alpha, n, {\bf R}_m \right\rangle
\left\langle \alpha, n, {\bf R}_m \right| = 1.
$$

In the framework of the effective mass approximation it is assumed 
that the envelope wave functions 
$\psi^\alpha_n({\bf k})= \left\langle \alpha, n, {\bf k} | \psi 
\right\rangle $ 
and matrix elements of the Hamiltonian (2) (excluding the short-range 
part) are localized in a small area of $\bf k$-space inside of the 
Brillouin zone and the summation in (3) can be extended to the 
whole infinite $\bf k$-space. In this approximation we can consider  
${\bf R}_m$ as a continuous variable  $\bf R$ and the 
pseudo-Wannier basis transforms into pseudo-coordinate basis

$$
\left| \alpha, n, {\bf R} \right\rangle = \frac {1}{\sqrt V} 
\sum\limits_{\bf k}
\left| \alpha, n, {\bf k} \right\rangle e^{-i{\bf k}{\bf R}}
, \eqno (4)
$$

\par \noindent 
which satisfies the following conditions of orthonormality and completeness:

$$
\left\langle \alpha, n, {\bf R} | \alpha^\prime, n^\prime, {\bf R}^\prime 
 \right\rangle = \delta_{\alpha  \alpha^\prime }\delta_{n n^\prime }
\delta({\bf R}-{\bf R}^\prime), 
$$

$$
\sum\limits_{\alpha, n} \int d{\bf R} \left| \alpha, n, {\bf R} \right\rangle
\left\langle \alpha, n, {\bf R} \right| = 1.
\eqno (5)
$$

\par \noindent 
The Hamiltonian (1) in this basis according to (2), (4) and (5) 
has the following form:

$$
\begin {array}{c}
\left\langle \alpha, n, {\bf R} | H | \alpha^\prime, n^\prime, {\bf R}^\prime 
 \right\rangle = 
\frac {1}{V} \sum\limits_{{\bf k},{\bf k}^\prime}
e^{i{\bf kR}} \left\langle \alpha, n, {\bf k} | H | 
\alpha^\prime, n^\prime, {\bf k}^\prime 
 \right\rangle e^{-i{\bf k}^\prime {\bf R}^\prime}=\\
\left[E^{(\alpha)} \delta_{\alpha  \alpha^\prime }+
{\tilde f}({\bf R}) \Delta U^{\alpha  \alpha^\prime } 
\right] \delta_{n n^\prime }
\delta({\bf R}-{\bf R}^\prime)- 
\frac{{\hbar}^2}{2m_0} \left[\nabla_{\bf R}^2 
\delta({\bf R}-{\bf R}^\prime) \right]
\delta_{\alpha  \alpha^\prime }  \delta_{n n^\prime }+\\
\frac{\hbar}{m_0} \left[-i \nabla_{\bf R} 
\delta({\bf R}-{\bf R}^\prime) \right]
{\bf p}^{\alpha  \alpha^\prime }_{ n n^\prime }+
V^{\alpha  \alpha^\prime }_{ n n^\prime }({\bf R}) 
\delta({\bf R}-{\bf R}^\prime).
\end {array}
\eqno (6)
$$

\par \noindent 
Here  ${\tilde f}({\bf R}) = \sum\limits_{\bf k} 
f({\bf k}) e^{i{\bf kR}} $ is a smooth 
modulating function, 
$ V^{\alpha  \alpha^\prime }_{ n n^\prime }({\bf R}) = 
\sum\limits_{\bf k} V^{\alpha  \alpha^\prime }_{ n n^\prime }
({\bf k}) e^{i{\bf kR}} $ 
are interface short range corrections. In the effective mass 
approximation we can 
consider  $ {\tilde f}({\bf R}) \approx \Theta ({\bf nR}) $ and 
$ V^{\alpha  \alpha^\prime }_{ n n^\prime }({\bf R}) \approx
{\tilde V}^{\alpha  \alpha^\prime }_{ n n^\prime }\delta({\bf nR}) $, where 
$\Theta({\bf nR})$ and $\delta({\bf nR})$ are one-dimensional step-function 
and $\delta$-function respectively. 

The short range part of the Hamiltonian, characterized by matrix 
${\tilde V}^{\alpha \alpha^\prime}_{n n^\prime}$, 
is determined by the nature of atoms constituting the heterointerface 
and their disposition.
For ideal  heterointerface the elements of this matrix can be, 
in principal, calculated by microscopic theory \cite{14} - \cite{17}. 
In the effective mass approximation symmetry
of the short range corrections is determined exclusively by the symmetry 
of the crystal and orientation of the interface. 
Therefore it can be included into 
the Hamiltonian as an additional phenomenological term. 
Near to the interface plane, which is determined by the equation 
${\bf nx}=0$, the short range part of the interface potential  
$V^S({\bf nx})$ can be 
written in the form of the following decomposition:

$$
V^S({\bf nx}) = V^S(0) + {V^S}^\prime(0) \sum\limits_i n_i x_i + 
\frac{1}{2}{V^S}^{\prime\prime }(0) 
\sum\limits_{i, j} n_i n_j x_i x_j + \cdots .
$$

\par \noindent 
The position of the interface is determined by the condition 
$ V^S(0)=0$. 
Therefore the first term in the right hand side of this decomposition 
can be omitted. The second term, which is proportional to 
${V^S}^\prime(0)$, 
characterizes magnitude of the uniform electric field localized in the 
interface region and is the most significant. The third and the following 
terms characterize non-uniformity of this field. 
Thus from the phenomenological standpoint the matrix of 
the short range interface correction $\tilde V$ 
can be written as a decomposition in the powers of $\bf n$

$$
\tilde V = \sum\limits_i {\tilde V}_i n_i + 
\frac{1}{2}\sum\limits_{i, j} {\tilde V}_{ij} n_i n_j + \cdots ,
$$

\par \noindent 
in which the most significant role plays the term  proportional to $\bf n$.

The locality of the Hamiltonian (6) allows us, at every point of 
$\bf R$-space, to get rid of the off-diagonal 
matrix elements  ${\tilde f}({\bf R})  \Delta U^{\alpha  \alpha^\prime }$ 
which mix states of the same symmetry. To do that let us
 introduce a unitary matrix ${\bf S}({\bf R})$, which at any 
point $\bf R$ satisfies the conditions: 
${\bf S}^+({\bf R}){\bf S}({\bf R}) = {\bf 1}$  
($\bf 1$ is a unit matrix), and diagonalizes smooth part of the potential 
energy operator of the Hamiltonian (6)

$$
\sum\limits_{{\alpha}^{\prime\prime},{\alpha}^{\prime \prime \prime }}
S^{+\alpha \alpha ^{\prime\prime }} ({\bf R})
\left[E^{(\alpha^{\prime \prime})} \delta_{\alpha^{\prime\prime} 
\alpha^{\prime\prime\prime} }+
{\tilde f}({\bf R}) \Delta U^{\alpha^{\prime \prime}  
\alpha^{\prime\prime\prime} } \right]
S^{  \alpha^{\prime\prime\prime}\alpha^\prime}({\bf R}) =
E^{(\alpha)} ({\bf R}) \delta_{\alpha \alpha^\prime}
 .\eqno (7)
$$

\par \noindent 
Here  $ E^{(\alpha)} ({\bf R})$ is position-dependent 
energy of electron in  $\alpha$ 
band at ${\bf k}=0$, which determines the band offset at the interface. 
Matrix elements $S^{\alpha \alpha ^\prime}({\bf R})$  are solutions of the 
following system of eigenvalue equations:

$$
\sum\limits_{\alpha^\prime}
\left[E^{(\alpha)} \delta_{\alpha \alpha^\prime}+
{\tilde f}({\bf R}) \Delta U^{\alpha \alpha^\prime} \right]
S^{  \alpha^\prime \alpha^{\prime\prime}  }({\bf R}) =
E^{(\alpha^{\prime\prime})} ({\bf R}) 
S^{\alpha \alpha^{\prime\prime }}({\bf R}),
$$

\par \noindent 
where $\bf R$ is a parameter.

Unitary transformation of the Hamiltonian (6) 
by the matrix  ${\bf S}({\bf R })$
 is equivalent to using a new position-dependent pseudo-coordinate basis 

$$
\left| S^{\alpha}_n , {\bf R}\right\rangle =
\sum\limits_{\alpha^\prime} \left| \alpha^\prime , n, {\bf R} \right\rangle
S^{  \alpha^\prime \alpha }({\bf R}),
$$

\par \noindent 
which in the effective mass approximation takes account of variation 
of the Bloch states at ${\bf k} = 0$ across interface. The 
Schr{\"o}dinger equation in this basis can be written as:

$$
\sum\limits_{\alpha^\prime, n^\prime} \int d{\bf R}^\prime
\left\langle S^{\alpha}_n , {\bf R}| H | S^{\alpha^\prime}_{n^\prime}
 , {\bf R}^\prime \right\rangle  
\left\langle S^{\alpha^\prime}_{n^\prime} , {\bf R}^\prime| 
\psi \right\rangle =
E \left\langle S^{\alpha}_n , {\bf R}| \psi \right\rangle
,\eqno (8)
$$

\par \noindent 
where the Hamiltonian is

$$
\left\langle S^{\alpha}_n , {\bf R}| H | S^{\alpha^\prime}_{n^\prime}
 , {\bf R}^\prime \right\rangle =
\sum\limits_{\alpha^{\prime\prime}, \alpha^{\prime \prime \prime }}
S^{+\alpha  \alpha^{\prime\prime }}({\bf R}) 
\left\langle \alpha^{\prime\prime }, n, {\bf R} \left| H 
\right| \alpha^{\prime\prime \prime }, n^\prime, {\bf R}^\prime \right\rangle
S^{ \alpha^{\prime\prime \prime} \alpha^{\prime}}({\bf R}^\prime)
.\eqno (9)
$$

\par \noindent 
Transformation (9) does not break the locality of the Hamiltonian 
of the heterosystem and therefore the equation (8) is a differential 
equation and can be represented as:

$$
\sum\limits_{\alpha^\prime, n^\prime}
\left[E^{(\alpha)} ({\bf R}) \delta_{\alpha \alpha^\prime}\delta_{n n^\prime}+
T ^{\alpha \alpha^\prime}_{n n^\prime}(\hat{\bf k} , {\bf R}) \right]
\psi^{  \alpha^\prime}_{n^\prime}({\bf R}) =
E \psi^\alpha_n({\bf R})
 ,\eqno (10)
$$

\par \noindent 
where 

$$
T ^{\alpha \alpha^\prime}_{n n^\prime}(\hat{\bf k} , {\bf R})=
\sum\limits_{{\alpha}^{\prime\prime},{\alpha}^{\prime \prime \prime }}
S^{+\alpha \alpha ^{\prime\prime}} ({\bf R})
\left[ \frac{\hbar^2}{2m_0} {\hat {\bf k}}^2
\delta_{\alpha^{\prime\prime} \alpha^{\prime\prime\prime} }
\delta_{n n^\prime}+ \frac{\hbar}{m_0} 
{\hat {\bf k}} {\bf p}^{\alpha^{\prime\prime} \alpha^{\prime\prime\prime} }
_{n n^\prime}+
V^{\alpha^{\prime\prime} \alpha^{\prime\prime\prime} }_{n n^\prime}
({\bf R}) \right] S^{\alpha^{\prime\prime \prime} \alpha^\prime }
({\bf R})
$$

\par\noindent and $\hat{\bf k}=-i\partial/\partial{\bf R}$, 
$\psi^\alpha_n({\bf R}) =   \left\langle S^{\alpha}_n , {\bf R}| 
\psi \right\rangle$.

Renormalized single-band Hamiltonian 
$H^{(\alpha)}_{n n^\prime}(\hat{\bf k} , {\bf R})$ 
is obtained by excluding from the system of equations (10) all states 
$\beta \ne \alpha$ \cite{4}. Green's function  
$G^{(\beta)}(E, {\bf R}) = 1/ \left[ E - E^{(\beta)}({\bf R}) \right]$, 
which is usually used in this procedure, can be approximately substituted by 
$G^{(\beta)} \left[E^{\alpha}({\bf R}), {\bf R}\right] =
 1/ \left[ E^{\alpha}({\bf R})  - E^{(\beta)}({\bf R}) \right]$. 
Up to the terms of the second order this corresponds to the standard 
perturbation theory. As a result we get

$$
\begin {array}{c}
H^{(\alpha)}_{n n^\prime}(\hat{\bf k} , {\bf R})=
E^{(\alpha)}({\bf R}) \delta_{n n^\prime}+
T ^{\alpha \alpha}_{n n^\prime}(\hat{\bf k} , {\bf R})+\\
\sum\limits_{{\beta \ne \alpha ,\atop m}}
T ^{\alpha \beta}_{n m}(\hat{\bf k} , {\bf R})
\frac{1}{ E^{(\alpha)}({\bf R})  - E^{(\beta)}({\bf R}) }
T ^{ \beta \alpha }_{m n^\prime}(\hat{\bf k} , {\bf R}) +
\cdots, 
\end {array}
\eqno (11)
$$

\par \noindent 
where omitted right hand side terms correspond to 
the corrections of the third 
and higher orders of the perturbation theory. 

Commutation relations 

$$
{\bf S}^+({\bf R }){\hat{\bf k}}- {\hat{\bf k}}{\bf S}^+({\bf R })
=i\partial {\bf S}^+({\bf R })/\partial{\bf R},
$$

$$
{\bf S}({\bf R }){\hat{\bf k}}- {\hat{\bf k}}{\bf S}({\bf R })
=i\partial{\bf S}({\bf R })/\partial{\bf R},
$$

\par \noindent 
allow us to transform the single-band $\bf kp$-Hamiltonian (11)
to the canonical form. The additional terms, containing 
$\partial {\bf S}^+({\bf R })/ \partial{\bf R}$ 
and  
$\partial{\bf S}({\bf R })/ \partial{\bf R} \sim {\bf n} \delta ({\bf nR})$, 
arise due to difference of the Bloch wave functions on 
the both sides of the interface and in the framework of the effective mass 
approximation can be attached to the short range $\delta$-corrections.

Taking into account all stated above the single-band Hamiltonian (11)
 can be written as a sum of two terms

$$
H^{(\alpha)}_{n n^\prime}(\hat{\bf k} , {\bf R})=
H^{0(\alpha)}_{n n^\prime}(\hat{\bf k} , {\bf R})+
H^{S(\alpha)}_{n n^\prime}(\hat{\bf k} , {\bf R})
. \eqno (12)
$$

\par \noindent 
Here  
$ H^{0(\alpha)}_{n n^\prime}(\hat{\bf k} , {\bf R})$
is a bulk-like part of $\bf kp$-Hamiltonian which can be written up 
to the terms proportional to ${\hat k}_i {\hat k}_j$ in the following form:

$$ H^{0(\alpha)}_{n n^\prime}(\hat{\bf k} , {\bf R}) =
E^{(\alpha)} ({\bf R}) \delta_{n n^\prime}+
\frac {1}{2} \frac {\hbar}{m_0}
\left[ \hat {\bf k} {\bf p}^{(\alpha)}_{n n^\prime} ({\bf R})+
{\bf p}^{(\alpha)}_{n n^\prime} ({\bf R}) \hat {\bf k}
\right]+ \frac {\hbar^2}{2m_0}{\hat k}_i 
M^{(\alpha)}_{{n n^\prime} \atop{ ij}}
({\bf R}) {\hat k}_j
,\eqno  (13)
$$

\par \noindent 
where 

$$
M^{(\alpha)}_{{n n^\prime}\atop {ij}}({\bf R}) =
\delta_{ij} \delta_{n n^\prime}+
\frac {2}{m_0} \sum\limits_{{\beta \ne \alpha,} \atop m}
\frac{{\left[{\bf p}^{\alpha \beta}_{n m} ({\bf R})\right]}_i
{\left[{\bf p}^{ \beta \alpha }_{m n^\prime} ({\bf R}) \right]}_j}
{ E^{(\alpha)}({\bf R})  - E^{(\beta)}({\bf R}) }
\eqno (14)
$$

\par \noindent 
is a matrix determining position-dependent band parameters $a({\bf R})$,

$$
{\bf p}^{ \alpha \alpha^\prime }_{n n^\prime} ({\bf R}) =
\sum\limits_{ \alpha^{\prime \prime } , \alpha^{\prime \prime \prime }}
S^{+ \alpha \alpha^{\prime\prime} } ({\bf R}) 
{\bf p}^{ \alpha ^{\prime \prime }\alpha^{\prime\prime\prime} }_{n n^\prime}
S^{ \alpha^{\prime \prime \prime } \alpha^\prime} ({\bf R})
\eqno (15)
$$

\par \noindent 
are position-dependent momentum operator matrix elements, 
${\bf p}^{ (\alpha )}_{n n^\prime} ({\bf R}) =
{\bf p}^{ \alpha \alpha}_{n n^\prime} ({\bf R}) $. 
In the effective-mass approximation all position-dependent parameters in 
the bulk-like part of the Hamiltonian are proportional to the step-function 
$\Theta ({\bf nR})$. 
The second term 
$ H^{S(\alpha)}_{n n^\prime}(\hat{\bf k} , {\bf R})$, which vanishes in the 
bulk material, 
is a short range part of the $\bf kp$-Hamiltonian. 
It contains interface short range $\delta$-corrections, 
including terms proportional to 
$n_i$, $n_i n_j$, $n_i {\hat k}_j$, $n_i n_j {\hat k}_l$, $\ldots$ . 
In the effective mass approximation  
$\delta^2 ({\bf nR}) \approx (1/a) \delta ({\bf nR})$,
$\delta^3 ({\bf nR}) \approx (1/a^2) \delta ({\bf nR})$,
$\ldots$, 
where $a$ is a characteristic size of the interface region, 
and all parameters, which determine the short-range part of 
the Hamiltonian, are position-independent.  In this approximation

$$
\begin {array} {c}
H^{S(\alpha)}_{n n^\prime}(\hat{\bf k} , {\bf R}) =
\Biggl[\sum\limits_i A^i _{n n^\prime}n_i \delta ({\bf nR}) +
\sum\limits_{i,j} B^{ij} _{n n^\prime}n_i n_j \delta ({\bf nR}) +
\sum\limits_{i,j} C^{ij} _{n n^\prime}n_i {\hat k}_j \delta ({\bf nR}) + \\
\sum\limits_{i,j,l} D^{ijl} _{n n^\prime}{\hat k}_i {\hat k}_j n_l 
\delta ({\bf nR}) +
\sum\limits_{i,j,l} K^{ijl} _{n n^\prime}{\hat k}_i n_l 
\delta ({\bf nR}) {\hat k}_j + \cdots\Biggr] + H.C.
\end {array}
$$

\par \noindent 
All these terms can be introduced phenomenologically 
by the method of invariants, 
based only on the symmetry of the states under 
consideration (see Section III).

To take into account spin-orbit interaction it is necessary to add to 
the Hamiltonian (1) an operator $H_{so}$, which determines spin-orbit 
coupling. For heterosystems this operator includes three 
terms which mix states 
with different spin projections

$$
\begin {array}{c}
H_{so}= \frac{\hbar}{4 m^2_0 c^2}
\left( \left[ \nabla U_1({\bf x}) {\bf p}\right] \cdot {\bf \sigma} \right)+
\frac {\hbar}{4 m^2_0 c^2} f ({\bf x})
\left( \left[ \nabla \left(\Delta U({\bf x}) \right) {\bf p}\right] 
\cdot {\bf \sigma} \right)+ \\
\frac {\hbar}{4 m^2_0 c^2} \Delta U ({\bf x})
\left( \left[ \nabla f({\bf x})  {\bf p}\right] \cdot {\bf \sigma} \right)
\end {array}
$$

\par \noindent
The first term in the right hand side of this expression is the 
usual operator of spin-orbit interaction in the bulk crystal with periodic potential 
$ U_1({\bf x}) $.
 The second term determines the change of this operator across the 
interface. The third term is non-zero only in a narrow 
interface region and characterizes spin-dependent 
interface short range corrections.

Analysis shows that the spin-orbit coupling does not lead to a 
qualitative change of the effective Hamiltonian (12). 
The dependence  
$E^{(\alpha)}({\bf R})$ in (13) still determines the interface band 
offsets, but takes account of the position-dependent spin-orbit 
splitting. The matrix elements of momentum operator $\bf p$ in (13)-(15) 
must be substituted by matrix elements of operator 
$\bf \pi$, which in the case 
of a heterointerface system has the following form

$$
\pi= {\bf p}+ \frac {\hbar}{4 m^2_0 c^2}
\left[ {\bf \sigma} \nabla U_1({\bf x}) \right] +
\frac {\hbar}{4 m^2_0 c^2} f ({\bf x})
\left[{\bf \sigma} \nabla \left(\Delta U({\bf x}) \right) \right].
$$

\par \noindent 
All short range corrections, that arise due to the spin-orbit coupling, 
are relativistic and in the framework of the effective mass approximation 
can be included into the short range part of the Hamiltonian (12).

\begin {center}
\bf
III. $\bf kp$-HAMILTONIAN IN HETEROINTERFACE SYSTEM. 
METHOD OF INVARIANTS
\end {center}

The effective Hamiltonian (12), containing real 
position-dependent band parameters 
$a({\bf R})$, can be obtained by the method of invariants. To do that it is 
necessary to perform preliminary transformation to $\bf k$-representation. 
In this 
representation the matrix of the effective $\bf kp$-Hamiltonian 
depends on the 
three variables $\bf k$,  ${\bf k}^\prime$ and $\bf n$, and includes
Fourier transformants of the band parameters $a({\bf k}-{\bf k}^\prime)$ 
as "constants" of the method of invariants. 
This matrix must satisfy the invariance condition

$$
\sum\limits_{ n^{\prime \prime } , n^{\prime \prime \prime }}
D^{(\alpha)}_{ n^ \prime  n^{ \prime \prime }}(g)
H^{(\alpha)} _{ n^{\prime \prime } n^{\prime \prime \prime }}
(g^{-1}{\bf k}, g^{-1}{\bf k}^\prime; g^{-1}{\bf n})
D^{(\alpha)^+}_{ n^{\prime \prime \prime } n^ \prime  }(g)=
H^{(\alpha)}_{ n n^\prime }({\bf k},{\bf k}^\prime; {\bf n} )
,\eqno (16)
$$

\par \noindent 
where $D^{(\alpha)}(g)$ are the $\Gamma_\alpha$ 
irreducible representation matrices 
of the crystal point-group $F (g \in F)$. The matrix of the Hamiltonian 
must be also Hermitian 

$$
H^{(\alpha)^*} _{ n n^\prime }({\bf k},{\bf k}^\prime; {\bf n} )=
H^{(\alpha)} _{ n^\prime n}({\bf k}^\prime, {\bf k}; {\bf n} )
\eqno(17)
$$

\par \noindent 
and besides that it should satisfy the additional conditions of 
time-inversion invariance

$$
H^{(\alpha^*)} _{ n n^\prime }(-{\bf k},-{\bf k}^\prime; {\bf n} )=
H^{(\alpha)^*} _{ n n^\prime }({\bf k},{\bf k}^\prime; {\bf n} ).
\eqno (18)
$$

\par \noindent 
Here 
$\left| \alpha^*, n \right\rangle = T \left| \alpha, n \right\rangle $, 
where $T$ is the time-inversion operator.

Since the band parameters $a({\bf R})$ change only in the perpendicular 
to the interface direction, i.e. they are functions only of 
$ R_\bot = {\bf nR}$, their Fourier transformants are 

$$
\begin {array}{c}
a({\bf k}-{\bf k}^\prime) = \frac{1}{V}
\int a({\bf R}) e^{-i({\bf k}-{\bf k}^\prime) {\bf R}} d{\bf R} = \\
\frac{1}{V}
\int a( R_\bot) e^{-i(k_\bot- k^\prime_\bot)
 R_\bot } d R_\bot
e^{-i({\bf k}_{\|}-{\bf k}^\prime_{\|}) {\bf R}_{\|}} d{\bf R}_{\|}=
a(k_\bot- k^\prime_\bot)
\delta_{{\bf k}_{\|}{\bf k}^\prime_{\|}},
\end {array}
$$

\par \noindent 
where ${\bf R}_{\|} ={\bf R} - {\bf n} R_{\bot}$. 
These parameters are invariants under the symmetry 
transformation (16) and also satisfy the additional 
conditions (17) and (18).

The method of invariants allows one to obtain the matrix of the 
effective $\bf kp$-Hamiltonian of the heterointerface system in 
any order of the perturbation theory  in $\bf k$ and $\bf n$ with arbitrary 
"constants" (band parameters) $a({\bf k}-{\bf k}^\prime)$.

We are mostly interested in $A_3B_5$-heterosystems (point-group $T_d$). 
Table I. contains Hermitian combinations of $\bf k$, ${\bf k}^\prime$ and 
$\bf n$, transforming according to irreducible representations of 
this group, up to the terms proportional $kn^2$.

{\footnotesize
\begin{tabular}{|c|c|c|}
\multicolumn {3}{p{13cm}}{
TABLE I. Hermitian combinations of $\bf k$, ${\bf k}^\prime$ and 
$\bf n$, transforming according to the irreducible representations 
of the point-group $T_d$.}\\
\hline
\hline
Representation 
& 
Even under time inversion${}^a$
& 
Odd under time inversion${}^a$
\\
\hline
\hline 
$\Gamma_{15}$ 
& 
$i({\bf k}-{\bf k}^\prime)$ 
& 
$({\bf k}+{\bf k}^\prime)$
\\ 
\hline
$\Gamma_1$ 
& 
$({\bf k} \cdot {\bf k}^\prime)$ 
& 
\\
\hline
$\Gamma_{12}$ 
& 
$
\left\{
\begin {array} {c}
2 k_z k^\prime_z - k_x k^\prime_x- k_y k^\prime_y \\
\sqrt 3 (k_x k^\prime_x- k_y k^\prime_y)
\end {array}
\right.
$
& 
\\
\hline
$\Gamma_{15}$
&
$
\left\{ k_y k^\prime_z \right\}, 
\left\{ k_z k^\prime_x \right\}, 
\left\{ k_x k^\prime_y \right\}
$
&
\\
\hline
$\Gamma_{25}$
&
&
$i\left[{\bf k} \times {\bf k}^\prime \right]$
\\
\hline 
$\Gamma_{15}$
&
$\bf n$
&
\\
\hline 
$\Gamma_1$
&
$({\bf n} \cdot {\bf n}) = 1$
&
\\
\hline 
$\Gamma_{12}$
&
$
\left\{
\begin {array} {c}
2n^2_z- n^2_x - n^2_y \\
\sqrt 3 \left( n^2_x - n^2_y \right)
\end {array}
\right.
$
&
\\
\hline 
$\Gamma_{15}$
&
$
n_y n_z, 
n_z n_x, 
n_x n_y 
$
&
\\
\hline 
$\Gamma_1$
&
$i({\bf k}-{\bf k}^\prime) \cdot {\bf n}$
&
$({\bf k}+{\bf k}^\prime)  \cdot {\bf n}$
\\
\hline 
$\Gamma_{12}$
&
$
\left\{ 
\begin {array}{c}
i\Big[2(k_z -k^\prime_z)n_z - (k_x -k^\prime_x)n_x \\
- (k_y- k^\prime_y)n_y\Big]\\
i\sqrt 3\left[ (k_x -k^\prime_x)n_x- (k_y -k^\prime_y)n_y \right]
\end {array}
\right.
$
& 
$
\left\{ 
\begin {array}{c}
2(k_z  +k^\prime_z)n_z - (k_x +k^\prime_x)n_x \\
- (k_y +k^\prime_y)n_y\\
\sqrt 3\left[ (k_x +k^\prime_x)n_x- (k_y +k^\prime_y)n_y \right]
\end {array}
\right.
$
\\
\hline 
$\Gamma_{15}$
&
$
\left\{ 
\begin {array} {c}
i\left\{ (k_y -k^\prime_y)n_z \right\}\\
i\left\{ (k_z -k^\prime_z)n_x \right\}\\
i\left\{ (k_x -k^\prime_x)n_y \right\}
\end {array}
\right.
$
&
$
\left\{
\begin {array} {c}
\left\{ (k_y +k^\prime_y)n_z \right\}\\
\left\{ (k_z +k^\prime_z)n_x \right\}\\
\left\{ (k_x +k^\prime_x)n_y \right\}
\end {array}
\right.
$
\\
\hline 
$\Gamma_{25}$
&
$i\left[({\bf k}-{\bf k}^\prime)\times {\bf n}\right]$
&
$\left[({\bf k}+{\bf k}^\prime)\times {\bf n}\right]$
\\
\hline 
$\Gamma_{15}$
&
$i({\bf k}-{\bf k}^\prime)\cdot {\bf n}^2 = i({\bf k}-{\bf k}^\prime)$
&
$({\bf k}+{\bf k}^\prime)\cdot {\bf n}^2 = ({\bf k}+{\bf k}^\prime)$
\\
\hline 
$\Gamma_{15}$
&
$
\left\{
\begin {array} {c}
i(k_x -k^\prime_x)(2n^2_x- n^2_y- n^2_z)\\
i(k_y -k^\prime_y)(2n^2_y- n^2_z- n^2_x)\\
i(k_z -k^\prime_z)(2n^2_z- n^2_x- n^2_y)
\end {array}
\right.
$
&
$
\left\{
\begin {array} {c}
(k_x +k^\prime_x)(2n^2_x- n^2_y- n^2_z)\\
(k_y +k^\prime_y)(2n^2_y- n^2_z- n^2_x)\\
(k_z +k^\prime_z)(2n^2_z- n^2_x- n^2_y)
\end {array}
\right.
$
\\
\hline 
$\Gamma_{25}$
&
$
\left\{
\begin {array} {c}
i(k_x -k^\prime_x)(n^2_y- n^2_z)\\
i(k_y -k^\prime_y)(n^2_z- n^2_x)\\
i(k_z -k^\prime_z)( n^2_x- n^2_y)
\end {array}
\right.
$
&
$
\left\{
\begin {array} {c}
(k_x +k^\prime_x)(n^2_y- n^2_z)\\
(k_y +k^\prime_y)(n^2_z- n^2_x)\\
(k_z +k^\prime_z)( n^2_x- n^2_y)
\end {array}
\right.
$
\\
\hline 
$\Gamma_1$
&
$i({\bf k}-{\bf k}^\prime) \cdot {\bf N}$
&
$({\bf k}+{\bf k}^\prime) \cdot {\bf N}$
\\
\hline 
$\Gamma_{12}$
&
$
\left\{
\begin {array} {c}
i\Bigl[2(k_z -k^\prime_z)N_z - (k_x -k^\prime_x)N_x \\
- (k_y- k^\prime_y)N_y\Bigr]\\
i\sqrt 3\left[ (k_x -k^\prime_x)N_x- (k_y -k^\prime_y)N_y \right]
\end {array}
\right.
$
& 
$
\left\{
\begin {array} {c}
2(k_z +k^\prime_z)N_z - (k_x +k^\prime_x)N_x \\
- (k_y+k^\prime_y)N_y\\
\sqrt 3 \left[ (k_x +k^\prime_x) N_x- (k_y +k^\prime_y)N_y \right]
\end {array}
\right.
$
\\
\hline 
$\Gamma_{15}$
&
$
\left\{
\begin {array} {c}
i \left\{ (k_y -k^\prime_y)N_z \right\}\\
i \left\{ (k_z -k^\prime_z)N_x \right\}\\
i \left\{ (k_x -k^\prime_x)N_y \right\}
\end {array}
\right.
$
&
$
\left\{
\begin {array} {c}
\left\{ (k_y +k^\prime_y)N_z \right\}\\
\left\{ (k_z +k^\prime_z)N_x \right\}\\
\left\{ (k_x +k^\prime_x)N_y \right\}
\end {array} 
\right.
$
\\
\hline 
$\Gamma_{25}$
&
$i\left[({\bf k}-{\bf k}^\prime)\times {\bf N}\right]$
&
$\left[({\bf k}+{\bf k}^\prime)\times {\bf N}\right]$
\\
\hline
\hline 
\multicolumn {3} {l} {${}^a$$N_x = n_y n_z$, 
$N_y = n_z n_x$, $N_z = n_x n_y$,
$\{A_i B_j \} = 1/2 (A_i B_j + A_j B_i)$.}
\end{tabular}
}

Subsequent transformation to $\bf R$-representation is performed by 
the following rules:

$$
a({\bf k}-{\bf k}^\prime) ({\bf k}\pm{\bf k}^\prime) \longrightarrow
{\hat {\bf k}} a({\bf nR})\pm a({\bf nR}) {\hat{\bf k}},
$$

$$a({\bf k}-{\bf k}^\prime) ( k_i  k^\prime_j) \longrightarrow 
{\hat  k}_i a({\bf nR}) {\hat k}_j.
$$

\par \noindent 
All the terms proportional to $n_i$ and $n_i n_j$ in $\bf R$-representation 
contain $\delta$-functions:

$$ 
n_i \longrightarrow n_i \delta({\bf nR}),
$$

$$ 
n_i n_j \longrightarrow n_i n_j \delta({\bf nR}),
$$

$$
(k_i \pm k^\prime_i) n_j \longrightarrow \left[{\hat  k}_i 
\delta({\bf nR}) \pm
\delta({\bf nR}){\hat  k}_i \right] n_j, 
$$

$$ 
(k_i \pm k^\prime_i) n_j n_l \longrightarrow \left[{\hat  k}_i 
\delta({\bf nR}) \pm
\delta({\bf nR}){\hat  k}_i \right] n_j n_l .
$$

\begin {center}
\bf
IV. EFFECTIVE HAMILTONIANS IN $A_3B_5$-HETEROSYSTEMS
\end {center}

As it was mentioned above a single band Hamiltonian in a 
heterointerface system can be represented as a sum of 
bulk-like and short-range parts:

$$
H^{(\alpha)} (\hat{\bf k} , {\bf R})=
H^{0(\alpha)}(\hat{\bf k} , {\bf R})+
H^{S(\alpha)}(\hat{\bf k} , {\bf R}).
$$

\par \noindent 
The short range part $ H^{S(\alpha)}(\hat{\bf k} , {\bf R})$ 
contains terms proportional to $n_i \delta({\bf nR})$, 
$n_i n_j \delta({\bf nR})$, $n_i {\hat k}_j \delta({\bf nR})$, 
$n_i {\hat k}_j \delta({\bf nR}) {\hat k}_l$, 
$\ldots$, which arise in the different orders of perturbation 
theory (11). These terms are interface short range corrections which 
change boundary conditions for smooth envelope functions in 
$\bf R$-representation. Analysis shows that in the short range part of 
the Hamiltonian it is sufficient to take into account only terms proportional 
to $n$, $n^2$, $kn$, $kn^2$, since these corrections can provide
proper form of subbands dispersion in quantum well, which is dictated
by symmetry of the quantum well (e.g. point-group $D_{2d}$ for a 
symmetric (001) quantum well). In this approximation the 
short range part of the Hamiltonian can be represented 
as a sum of four terms

$$
H^{S(\alpha)} (\hat{\bf k} , {\bf R})=
H^{S(\alpha)}_n ({\bf R})+
H^{S(\alpha)}_{n^2} ({\bf R})+
H^{S(\alpha)}_{kn}( \hat{\bf k} , {\bf R})+
H^{S(\alpha)}_{kn^2}(\hat{\bf k} , {\bf R}).
$$

In this Section the effective $\bf kp$-Hamiltonians of the
$\Gamma_1$, $\Gamma_6$ and $\Gamma_{15}$, $\Gamma_8$ bands, 
and also $\Gamma_8 {\oplus} \Gamma_ 7$ two-band Hamiltonian are 
presented. In the bulk-like part $ H^{0(\alpha)} (\hat{\bf k} , {\bf R})$ 
we considered $E^{(\alpha)} ({\bf R}) = 0$ 
and took into account only terms up to quadratic-in-$\bf k$. 
In the short range part we included only one term 
$ H^{S(\alpha)}_n ({\bf R})$, proportional to the first order of 
$\bf n$, since it is the strongest short range correction, characterizing 
a uniform electric field localized in the interface region. The rest of the short range 
corrections are too cumbersome and therefore given in Appendix A. 

It is necessary to note that the bulk-like part of the Hamiltonian  
$ H^{0(\alpha)} (\hat{\bf k} , {\bf R})$ includes position-dependent 
parameters, even though their explicit dependence on $\bf R$ is omitted. 
Parameters that describe the short range part of the Hamiltonian  
$ H^{S(\alpha)} (\hat{\bf k} , {\bf R})$ are position-independent.

\begin {center}
\bf
A. $\Gamma_1$ band  Hamiltonian
\end {center}

The $\Gamma_1$ conduction band  Hamiltonian is determined by one 
position-dependent parameter (inverse effective mass) 
$a = m_0/m^*_e$:

$$
H^{0(\Gamma_1)} (\hat{\bf k} , {\bf R})=
\frac{\hbar^2}{2m_0}
\left(\hat{\bf k} a \hat{\bf k} \right)
.$$

\par \noindent 
Here 
$\left(\hat{\bf k} a \hat{\bf k} \right) =
{\hat k}_x a {\hat k}_x+
{\hat k}_y a {\hat k}_y+
{\hat k}_z a {\hat k}_z$, 
$a = 1 + \frac{2}{m_0} \sum\limits_q
\frac{{\left| \left\langle \Gamma_{15}; q \left\|
{\bf p} \right\| \Gamma_1 \right\rangle \right|}^2}
{E^{\Gamma_1}-E^{\Gamma_{15}}_q}$.

The short range terms, proportional to the first order of $\bf n$, 
in the $\Gamma_1$ band Hamiltonian do not exist 
($H^{S(\Gamma_1)}_n ({\bf R}) = 0$).

\begin {center}
\bf
B. $\Gamma_6$ band Hamiltonian
\end {center}

The $\Gamma_6$ conduction band Hamiltonian is determined by two 
position-dependent parameters $\alpha_1$ and $\alpha_2$

$$
H^{0(\Gamma_6)} (\hat{\bf k} , {\bf R})=
\frac{\hbar^2}{2m_0}
\Biggl[ \left( \hat{\bf k} \alpha_1 \hat{\bf k} \right) I
+ i \left( \left[\hat{\bf k} \alpha_2 \hat{\bf k} \right] \cdot {\bf \sigma}
\right) \Biggr] .
$$

\par \noindent 
Here $I$ is a unit matrix $2\times2$, $\sigma_i$ are Pauli matrices, 
${\left[\hat{\bf k} a \hat{\bf k} \right]}_i =
{\hat k}_{i+1} a {\hat k}_{i+2}-{\hat k}_{i+2} a {\hat k}_{i+1} $.

The parameter $\alpha_1= m_0/m^*_e$ determines the electron inverse 
effective mass in the $\Gamma_6$ band and is equal to the parameter $a$ 
of the $\Gamma_1$ band, when we neglect spin-orbit coupling. 
The parameter $\alpha_2$ determines an additional relativistic term, 
which arises when the presence of heterointerface is taken into account. 
It vanishes in the bulk material due to commutativity of $\alpha_2$ and 
operator $\hat {\bf k}$. This term describes electron effective spin-orbit 
coupling at the heterointerface, which is calculated by means of 
envelope functions (interface effective spin-orbit interaction). 
Indeed, using the commutation relation  
$\alpha_2({\bf R})\hat{\bf k}-\hat{\bf k}\alpha_2({\bf R})=
i\partial \alpha_2({\bf R})/\partial {\bf R}$
and taking into account the fact that in the effective mass approximation 
$\partial \alpha_2({\bf R})/\partial {\bf R} \approx 
\Delta \alpha_2 {\bf n} \delta({\bf nR})$, 
this term can be represented in the form
$\Delta \alpha_2 \left( \left[{\bf n} \times \hat{\bf k} \right]
\cdot {\bf \sigma} \right)  \delta({\bf nR})$. 
It is easy to see that it is analogous to the operator of spin-orbit 
interaction, in which the factor proportional to 
$\Delta \alpha_2 {\bf n} \delta ({\bf nR})$ acts as an effective 
electric field localized in the interface region and $\hat {\bf k}$ is the momentum 
operator in the effective mass approximation. 

The parameter  $\alpha_2$ can be approximately determined if we 
take into account $\bf kp$-interaction of the 
$\Gamma_6 (\Gamma^c_1)$ conduction 
band only with the $\Gamma_8 {\oplus} \Gamma_7 (\Gamma^v_{15})$  
valence band. In this 
approximation it is proportional to the spin-orbit splitting of the 
$\Gamma^v_{15}$  band and to the square of the matrix element 
$\left\langle \Gamma^v_{15} \left\|
{\bf p} \right\| \Gamma^c_1 \right\rangle$, 
which determines the electron effective mass in the 
$\Gamma _6(\Gamma^c_1)$ band. Thus it can be approximately represented as

$$
\alpha_2 = \left[ \Delta_{so}/(3E_g + 2 \Delta_{so})\right]
\left( 1 - m_0/m_e \right),
$$

\par \noindent 
where $E_g = E^{\Gamma_6}_c - E^{\Gamma_8}_v$, 
$ \Delta_{so} = E^{\Gamma_8}_v - E^{\Gamma_7}_v$  is the 
spin-orbit splitting of the $\Gamma^v_{15}$ band.

The short range part, proportional to the first order of $\bf n$, in the 
$\Gamma_6$ band Hamiltonian does not exist 
($H^{S(\Gamma_6)}_n ({\bf R}) = 0$).

\begin {center}
\bf
C. $\Gamma_{15}$ band Hamiltonian
\end {center}

The bulk-like part of the $\Gamma_{15}$ band hole Hamiltonian, 
containing four quadratic-in-$\bf k$ terms and one 
linear-in-$\bf k$ term, can be written in the form

$$
\begin {array}{c}
H^{0(\Gamma_{15})} (\hat{\bf k} , {\bf R})=
\frac{\hbar^2}{2m_0}
\Biggl[ \left( \hat{\bf k} b_1 \hat{\bf k} \right) I -
6\sum\limits_i \left[{\hat k}_i b_2 {\hat k}_i-
\frac{1}{3} \left( \hat{\bf k} b_2 \hat{\bf k} \right) \right] J^2_i -\\
12 \sum\limits_i \left\{{\hat k}_i b_3 {\hat k}_{i+1}\right\}
\left\{ J_i J_{i+1} \right\} +
i 3 \left( \left[\hat{\bf k} b_4 \hat{\bf k} \right] \cdot {\bf J}\right) +
\frac{i}{a_0} 2 \sqrt 3 \sum\limits_i \left[ {\hat k}_i d \right]
\left\{ J_{i+1} J_{i+2} \right\}
\Biggr] ,
\end {array}
$$

\par \noindent 
where $I$ is a unit matrix $3\times3$, $J_i$ are matrices of the
angular momentum $J = 1$, 
$\left\{J_i J_j \right\} = \frac{1}{2}
\left(J_i J_j + J_j J_i \right)$, 
$\left\{{\hat k}_i a {\hat k}_j \right\} = \frac{1}{2}
\left({\hat k}_i a {\hat k}_j + {\hat k}_j a {\hat k}_i \right)$, 
$\left[ {\hat k}_i d \right]= {\hat k}_i d - d {\hat k}_i$, 
$a_0$ is the lattice constant, which was introduced for convenience.

Band parameters $b_l$ can be expressed by position-dependent parameters 
$\rho_l$, which describe $\bf kp$-interaction of the $\Gamma_{15}$ valence 
band with the $\Gamma_1$, 
$\Gamma_{12}$, $\Gamma_{15}$ and $\Gamma_{25}$  bands:

$$
b_1= -1 + 2 \rho_1+4\rho_2+4\rho_3+4\rho_4,
$$ 

$$
b_2=\rho_1-\rho_2-\rho_3+2\rho_4,
$$
 
$$
b_3=\rho_1-\rho_2+\rho_3-\rho_4,
$$

$$
b_4=\rho_1+\rho_2-\rho_3-\rho_4.
$$

\par \noindent 
Here

$$
\rho_1 = \frac{1}{3m_0} \sum\limits_q
\frac{{\left| \left\langle \Gamma_{15} \left\|
{\bf p} \right\| \Gamma_1; q \right\rangle \right|}^2}
{E^{\Gamma_1}_q-E^{\Gamma_{15}}},
$$

$$
\rho_2 = \frac{1}{6m_0} \sum\limits_q
\frac{{\left| \left\langle \Gamma_{15} \left\|
{\bf p} \right\| \Gamma_{25}; q \right\rangle \right|}^2}
{E^{\Gamma_{25}}_q-E^{\Gamma_{15}}},
$$

$$
\rho_3 =  \frac{1}{6m_0} \sum\limits_q
\frac{{\left| \left\langle \Gamma_{15} \left\|
{\bf p} \right\| \Gamma_{15}; q \right\rangle \right|}^2}
{E^{\Gamma_{15}}_q-E^{\Gamma_{15}}},
$$
 
$$
\rho_4 = \frac{1}{6m_0} \sum\limits_q
\frac{{\left| \left\langle \Gamma_{15} \left\|
{\bf p} \right\| \Gamma_{12}; q \right\rangle \right|}^2}
{E^{\Gamma_{12}}_q-E^{\Gamma_{15}}}.
$$

The position-dependent parameters $b_1$, $b_2$ and $b_3$ are 
analogous to the $\Gamma_{15}$ band parameters of the bulk 
material. The parameter $b_4$ arises due to the presence of the 
heterointerface and similarly to the parameter $\alpha_2$ of the
$\Gamma_6$ band describes interface effective spin-orbit interaction. 
In this case hole in the 
$\Gamma_{15}$ band should be attributed an effective spin 
$J = 1$. If $\bf kp$-interaction of the $\Gamma_{15}$ valence band 
with all the $\Gamma_{25}$ bands is neglected ($\rho_2 = 0$) 
the parameter $b_4$ can be expressed by all other parameters
$b_1$, $b_2$ and $b_3$.  In this approximation 
$b_4 = (1/3)(-1-b_1+2b_2+3b_3)$.

The additional linear-in-$\bf k$ term, which is proportional to a 
position-dependent parameter $d$, appears in the $\Gamma_{15}$  
band $\bf kp$-Hamiltonian only when the presence of interface 
is taken into account and arises due to variation of the Bloch functions 
across the interface. Indeed, the diagonal matrix elements of 
the momentum operator 
${\bf p}^{\Gamma_{15} \Gamma_{15}}_{n n^\prime}
={\bf p}^{(\Gamma_{15})}_{n n^\prime}$ 
are equal to zero since 
the operator $\bf p$ is odd under time inversion. However, this 
does not take place when the position-dependent pseudo-coordinate 
basis is used, since in this case, according to (15), the diagonal 
position-dependent 
matrix elements of the momentum operator also include 
off-diagonal matrix elements of the operator $\bf p$, which are non-zero 
in general

$$
{\bf p}^{(\Gamma_{15})}_{n n^\prime}({\bf R})=
\sum\limits_{\Gamma^\prime_{15} \Gamma^{\prime \prime}_{15}}
S^{+\Gamma_{15} \Gamma^\prime_{15}}({\bf R})
{\bf p}^{\Gamma^\prime_{15} \Gamma^{\prime \prime}_{15}}_{n n^\prime}
S^{\Gamma^{\prime\prime }_{15} \Gamma_{15}}({\bf R}).
$$

The short range part of the $\Gamma_{15}$ band Hamiltonian, which 
is proportional to the first order of $\bf n$, is determined by one 
position-independent parameter $r$ and has the form

$$
H^{S(\Gamma_{15})}_n ({\bf R})=
\frac{\hbar^2}{2m_0 a_0} 2 \sqrt 3 r
\sum\limits_i n_i
\left\{ J_{i+1} J_{i+2} \right\} \delta ({\bf nR}).$$

\par \noindent 
This short range correction coincides with the correction, 
introduced in Ref. \cite{14}. for the case of (001) interface, 
and generalizes it for arbitrary interface orientation.

\begin {center}
\bf
D. $\Gamma_8$ band Hamiltonian
\end {center}

The bulk-like part of the $\Gamma_8$ band hole Hamiltonian along with the 
usual quadratic-in-$\bf k$ terms, determined by position-dependent 
Luttinger parameters  $\gamma_1$, $\gamma_2$  and $\gamma_3$, 
is also characterized by two additional quadratic-in-$\bf k$ terms, 
determined by parameters  $\gamma_4$ and $\gamma_5$, and two 
linear-in-$\bf k$ terms, described by 
parameters $\beta_1$ and $\beta _2$ 

$$
\begin {array}{c}
H^{0(\Gamma_{8})} (\hat{\bf k} , {\bf R})=
\frac{\hbar^2}{2m_0}
\Biggl[ \left( \hat{\bf k} \gamma_1 \hat{\bf k} \right) I -
2\sum\limits_i \left[{\hat k}_i \gamma_2 {\hat k}_i-
\frac{1}{3} \left( \hat{\bf k} \gamma_2 \hat{\bf k} \right) \right] J^2_i - \\
4\sum\limits_i \left\{{\hat k}_i \gamma_3 {\hat k}_{i+1}\right\}
\left\{ J_i J_{i+1} \right\} +
i2 \left( \left[\hat{\bf k} \gamma_4 \hat{\bf k} \right] \cdot {\bf J}\right)+
i8 \sum\limits_i {\left[{\hat {\bf k}} 
\gamma_5 {\hat {\bf k}} \right]}_i J^3_i+\\
\frac{i}{a_0} \frac{2}{\sqrt 3}\sum\limits_i \left[ {\hat k}_i \beta_1 \right]
\left\{ J_{i+1} J_{i+2} \right\}+
\frac{1}{a_0}\frac{4}{\sqrt 3} \sum\limits_i \left\{ {\hat k}_i \beta_2 \right\}
\left\{J_i \left( J^2_{i+1} - J^2_{i+2}\right) \right\}
\Biggr] ,
\end {array}
$$

\par \noindent 
where $I$  is a unit matrix $4\times4$, $J_i$ are matrix of the angular 
momentum $J = 3/2$, 
$\left\{ {\hat k}_i a \right\}=
\frac{1}{2} \left({\hat k}_i a + a {\hat k}_i \right)$.

Neglecting spin-orbit interaction, the $\Gamma_8$ band parameters 
can be determined by the $\Gamma_{15}$ valence 
band parameters $b_l$:
$\gamma_1 = b_1$, $\gamma_2 = b_2$, 
$\gamma_3 = b_3$, $\gamma_4 = b_4$.  
The additional parameters $\gamma_4$ and $\gamma_5$, which describe 
quadratic-in-$\bf k$ terms, arise due to interface effective spin-orbit 
interaction, calculated by means of envelope functions (the hole effective 
spin $J = 3/2$). The parameter $\gamma_4$ is non-relativistic and can 
be determined approximately by the Luttinger parameters if 
$\bf kp$-interaction of the $\Gamma_{15}^v$ band with all the 
$\Gamma_{25}$ bands is neglected (i.e. $\rho_2 = 0$). In this case 
$\gamma_4 = (1/3)(-1-\gamma_1+2\gamma_2+3\gamma_3)$. The first four 
terms of the $\Gamma_8$ valence band  $\bf kp$-Hamiltonian, which are 
proportional to the parameters $\gamma_1$, $\gamma_2$, 
$\gamma_3$ and $\gamma_4$, coincide with  $\bf kp$-Hamiltonian obtained 
by Foreman in Ref. \cite{10}.
The parameter $\gamma_5$ is non-zero only if spin-orbit 
splitting of the intermediate states in the $\bf kp$-perturbation theory is 
taken into account. This parameter can be approximately determined, 
by taking into account $\bf kp$-interaction of the 
$\Gamma_8(\Gamma^v_{15})$ valence 
band only with the nearest
$\Gamma_7 {\oplus} \Gamma_8(\Gamma^c_{15})$ conduction band. Then it is 
$\gamma_5 = (1/18)(1+\gamma_1-2\gamma_2)
\left[ \Delta^\prime_{so}/(E^\prime_g + \Delta^\prime_{so}) \right]$, 
where $E^\prime_g = E_c^{\Gamma_7} - E_v^{\Gamma_8}$, 
$\Delta^\prime_{so} = E_c^{\Gamma_8} - E_c^{\Gamma_7}$
is the spin-orbit splitting of the $\Gamma^c_{15}$ conduction band.

The hole $\Gamma_8$ band Hamiltonian contains two linear-in-$\bf k$ 
terms, determined by the parameters $\beta_1$ and $\beta_2$. 
The parameter $\beta_1$, which arises due to variation of the Bloch 
functions across the interface, is non-relativistic and is originated from 
the corresponding parameter $d$ of the $\Gamma_{15}$ band 
($\beta_1= d$ if spin-orbit coupling is neglected). The parameter 
$\beta_2$ is relativistic and analogous to the corresponding parameter 
of the $\Gamma_8$  band in the bulk material, that arises due to the 
absence of the inversion symmetry in $A_3 B_5$-semiconductors. 

The short range part of the $\Gamma_8$  band Hamiltonian, which 
is proportional to the first order of $\bf n$, is determined by one 
parameter $\nu$, which is originated from the corresponding 
parameter $r$ of the $\Gamma_{15}$  band 
(when the spin-orbit coupling is neglected $\nu = r$) and has the form

$$
H^{S(\Gamma_{8})}_n ({\bf R})=
\frac{\hbar^2}{2m_0 a_0} \frac{2}{\sqrt 3} \nu
\sum\limits_i n_i
\left\{ J_{i+1} J_{i+2} \right\} \delta ({\bf nR}).
$$

\begin {center}
\bf
E. Two-band $\Gamma_8 {\oplus} \Gamma_7$ Hamiltonian
\end {center}

The $\Gamma_8 {\oplus} \Gamma_7$ two-band hole Hamiltonian
can be represented in the following matrix form:

$$
H^{(\Gamma_8 {\oplus} \Gamma_7)}({\hat {\bf k}}, {\bf R})=
\left[
\begin {array} {cc}
H^{(\Gamma_8)}({\hat {\bf k}}, {\bf R})&
H^{(\Gamma_8 \Gamma_7)}({\hat {\bf k}}, {\bf R})\\
 H^{(\Gamma_8 \Gamma_7)^+}({\hat {\bf k}}, {\bf R})& 
H^{(\Gamma_7)}({\hat {\bf k}}, {\bf R})
\end {array}
\right].
$$

\par \noindent 
Parameters that determine this Hamiltonian can be approximately 
related to corresponding parameters of the $\Gamma_8(\Gamma_{15}^v)$
band Hamiltonian if we neglect spin-orbit mixing. 
In this case $H^{(\Gamma_8)}({\hat {\bf k}}, {\bf R})$ coincides with the
$\Gamma_8$ band Hamiltonian (see Section IV.D), in which 
$\gamma_5 = \beta_2 = 0$. 
Note that the $\Gamma_7$ band Hamiltonian also includes a term 
which describe interface effective spin-orbit interaction. However 
in contrast to the  $\Gamma_6(\Gamma_1^c)$ band Hamiltonian this term is
 non-relativistic and proportional to 
the parameter $\gamma_4$. Also in this approximation 
the $H^{0(\Gamma_8 \Gamma_7)}({\hat {\bf k}}, {\bf R})$ block
contains one linear-in-$\bf k$ term proportional to the parameter 
$\beta_1$

$$
H^{0(\Gamma_7)} (\hat{\bf k} , {\bf R})=
\Delta_{so} I +
\frac{\hbar^2}{2m_0}
\Biggl[
\left( \hat{\bf k} \gamma_1 \hat{\bf k} \right) I
+ i2 \left( \left[\hat{\bf k} \gamma_4 \hat{\bf k} \right] \cdot {\bf \sigma}
\right)
 \Biggr] ,
$$

$$ 
\begin {array}{c}
H^{0(\Gamma_8 \Gamma_7)}({\hat {\bf k}}, {\bf R})=
\frac{\hbar^2}{2m_0}
\Biggl[
- \sqrt 2 \Bigl[ \left( 2{\hat k}_z \gamma_2 {\hat k}_z-
{\hat k}_x \gamma_2 {\hat k}_x-
{\hat k}_y \gamma_2 {\hat k}_y \right) I^{\Gamma_{12}}_1-\\
\sqrt 3 \left({\hat k}_x \gamma_2 {\hat k}_x-
{\hat k}_y \gamma_2 {\hat k}_y \right) I^{\Gamma_{12}}_2  \Bigr]-
\sqrt 6 \sum\limits_i \left\{ {\hat k}_{i+1} \gamma_3 {\hat k}_{i+2}\right\}
I^{\Gamma_{15}}_i +\\
i \frac{1}{\sqrt 2} \sum\limits_i \left[ {\hat k}_{i+1} \gamma_4 {\hat k}_{i+2}\right]
I^{\Gamma_{25}}_i +
\frac{i}{a_0} \frac{1}{\sqrt 2}\sum\limits_i \left[ {\hat k}_{i} \beta_1 \right]
I^{\Gamma_{15}}_i
\Biggr],
\end {array}
$$

\par \noindent 
where $I$ is a unit matrix $2 \times 2$, 
$\Delta_{so}=E^{\Gamma_8}_v- E^{\Gamma_7}_v $,
matrices $I^{\Gamma_{\alpha}}_i$ 
 are given in Appendix B.

The short range part of this Hamiltonian, which is 
proportional to the first order 
of $\bf n$, can be described by the corresponding parameters of the 
$\Gamma_8$ band

$$
H^{S(\Gamma_7)}_n ({\bf R}) = 0,
$$

$$
H^{S(\Gamma_8\Gamma_7)}_n ({\bf R})=
\frac{\hbar^2}{2m_0 a_0} \frac{1}{\sqrt 2} \nu
\sum\limits_i n_i I^{\Gamma_{15}}_i \delta ({\bf nR}).
$$

\begin {center}
\bf
V. BOUNDARY CONDITIONS
\end {center}

The boundary conditions for envelope functions can be 
obtained by integrating the effective-mass equation

$$
\sum\limits_{n^\prime} H^{(\alpha)}_{n n^\prime}(\hat{\bf k} , {\bf R})
\psi^{(\alpha)}_{n^\prime}({\bf R})=
E^{(\alpha)} \psi^{(\alpha)}_n ({\bf R}),
$$

\par  \noindent 
along the normal $\bf n$ across the interface. Taking into account in 
the effective-mass Hamiltonian 
$ H^{(\alpha)}(\hat{\bf k} , {\bf R})$ 
only bulk-like part 
$ H^{0(\alpha)}(\hat{\bf k} , {\bf R})$ 
and main short range term $ H^{(\alpha)}_n (\hat{\bf k} , {\bf R})$, 
proportional to the first order of $\bf n$, we obtain the following boundary 
conditions, that depend on the interface orientation

$$
\left. \psi \right|{}_{{\bf nR}=+0} = \left. \psi \right|{}_{{\bf nR}=-0}
,$$

$$
({\hat A}_{\bf nR} \psi ) \left|{}_{{\bf nR}=+0}\right. -
({\hat A}_{\bf nR} \psi ) \left|{}_{{\bf nR}=-0}\right.=
B_{\bf nR} \psi \left| {}_{{\bf nR}=0}\right.
,$$

\par  \noindent
here matrix operator ${\hat A}_{\bf nR}$ is determined by the 
bulk-like part $ H^{0(\alpha)}(\hat{\bf k} , {\bf R})$  
of the Hamiltonian, and matrix $ B_{\bf nR}$
 by its short range part $ H^{S(\alpha)}_n (\hat{\bf k} , {\bf R})$. 

The matrices ${\hat A}_{\bf nR}$  and $ B_{\bf nR}$
, that determine boundary conditions for all the bands discussed 
in this paper, are given below for the case of (001) interface, 
when $n_x = n_y = 0$, $n_z = 1$ (${\bf nR} = z$).

\begin {center}
\bf
A. $\Gamma_1$ band boundary conditions
\end {center}

$$
{\hat A}^{(\Gamma_1)}_z = 
\frac{\hbar^2}{2m_0}
\left(
a {\hat k}_z
\right),
$$

$$
 B^{(\Gamma_1)}_z =0.
$$

\begin {center}
\bf
B. $\Gamma_6$ band boundary conditions
\end {center}

$$
{\hat A}^{(\Gamma_6)}_z = 
\frac{\hbar^2}{2m_0}
\Biggl[ \left(
\alpha_1 {\hat k}_z \right) I +
i \alpha_2 \left( -\sigma_x {\hat k}_y +\sigma_y {\hat k}_x \right)
\Biggr],
$$

$$
 B^{(\Gamma_6)}_z =0.
$$

\begin {center}
\bf
C. $\Gamma_{15}$ band boundary conditions
\end {center}

$$
\begin {array} {c}
{\hat A}^{(\Gamma_{15})}_z = 
\frac{\hbar^2}{2m_0}
\Biggl[ \left(
b_1 {\hat k}_z \right) I -
2 b_2 \left( 2J^2_z- J^2_x- J^2_y \right) {\hat k}_z -\\
6 b_3 \left( \left\{J_y J_z\right\} {\hat k}_y
+\left\{J_z J_x \right\}{\hat k}_x \right) +
i3 b_4 \left( -J_x {\hat k}_y +J_y {\hat k}_x \right)+
\frac{i}{a_0} 2 \sqrt 3 d \left\{J_x J_y \right\}
\Biggr],
\end {array}
$$

$$
 B^{(\Gamma_{15})}_z =
- i \frac{\hbar^2}{2m_0 a_0} 2 \sqrt 3 r \left\{J_x J_y \right\}.
$$

\begin {center}
\bf
D. $\Gamma_8$ band boundary conditions
\end {center}

$$
\begin {array} {c}
{\hat A}^{(\Gamma_{8})}_z = 
\frac{\hbar^2}{2m_0}
\Biggl[ \left(
\gamma_1 {\hat k}_z \right) I -
\frac{2}{3} \gamma_2 \left( 2J^2_z- J^2_x- J^2_y \right) {\hat k}_z -\\
2 \gamma_3 \left( \left\{J_y J_z \right\} {\hat k}_y
+\left\{J_z J_x \right\}{\hat k}_x \right) +
i2\gamma_4 \left( -J_x {\hat k}_y +J_y {\hat k}_x \right)+
i8\gamma_5 \left( -J^3_x {\hat k}_y +J^3_y {\hat k}_x \right)+\\
\frac{i}{a_0} \frac{2}{\sqrt 3} \beta_1 \left\{J_x J_y \right\}+
\frac{1}{a_0} \frac{4}{\sqrt 3}\beta_2 
\left\{J_z \left( J^2_x - J^2_y \right) \right\}
\Biggr],
\end {array}
$$

$$
 B^{(\Gamma_{8})}_z =
- i \frac{\hbar^2}{2m_0 a_0}\frac{2}{\sqrt 3} 
\nu \left\{J_x J_y \right\}.
$$

\begin {center}
\bf
E. $\Gamma_8 {\oplus} \Gamma_7$ band boundary conditions
\end {center}

$$
{\hat A}^{(\Gamma_8 \oplus \Gamma_7)}_z = 
\left[
\begin {array}{cc}
{\hat A}^{(\Gamma_{8})}_z&{\hat A}^{(\Gamma_8 \Gamma_7)}_z\\
{\hat A}^{(\Gamma_8 \Gamma_7)^+}_z&{\hat A}^{(\Gamma_7)}_z
\end{array}
\right]
,$$

$$ 
B^{(\Gamma_8 \oplus \Gamma_7)}_z =
\left[
\begin {array}{cc}
B^{(\Gamma_{8})}_z&B^{(\Gamma_8 \Gamma_7)}_z\\
B^{(\Gamma_8 \Gamma_7)^+}_z&B^{(\Gamma_7)}_z
\end{array}
\right]
,  
$$

\par \noindent
where ${\hat A}^{(\Gamma_{8})}_z $ and 
$ B^{(\Gamma_{8})}_z$ coincide with corresponding matrices of the 
$\Gamma_8$ band (see Section V. D.), in which $\gamma_5 = 0$ and 
$\beta_2 = 0$.

$$
{\hat A}^{(\Gamma_7)}_z = 
\frac{\hbar^2}{2m_0}
\Biggl[ \left(
\gamma_1 {\hat k}_z \right) I +
i2 \gamma_4 \left( -\sigma_x {\hat k}_y +\sigma_y {\hat k}_x \right)
\Biggr],
$$

$$ 
\begin {array}{c}
{\hat A}^{(\Gamma_8 \Gamma_7)}_z=
\frac{\hbar^2}{2m_0}
\Biggl[
-2 \sqrt 2 \gamma_2 I^{\Gamma_{12}}_1 {\hat k}_z -
\sqrt 6 \gamma_3 \left( I^{\Gamma_{15}}_x {\hat k}_y+
I^{\Gamma_{15}}_y {\hat k}_x \right)+\\
i \frac{1}{\sqrt 2} \gamma_4 \left( -I^{\Gamma_{25}}_x {\hat k}_y+
I^{\Gamma_{25}}_y {\hat k}_x \right)+
\frac{i}{a_0}\frac{1}{\sqrt 2} \beta_1 I^{\Gamma_{15}}_z
\Biggr],
\end {array}$$

$$
 B^{(\Gamma_7)}_z =0,
$$

$$
 B^{(\Gamma_8 \Gamma_7)}_z =
-i \frac{\hbar^2}{2m_0 a_0} \frac{1}{\sqrt 2} \nu I^{\Gamma_{15}}_z.
$$

\par \noindent 
The $\bf k$-dependent short range terms, which were not included 
into the boundary conditions, along with $\delta$-functions contain 
also their derivatives, which modify the boundary conditions. 
However, influence of the omitted corrections is small and therefore when 
considering systems with heterointerface (e.g. quantum well) it is 
more convenient to take these corrections into account by the 
perturbation theory in the framework of the envelope function approximation.

\begin {center}
\bf
V. SUMMARY
\end {center}

In this paper using a position-dependent pseudo-coordinate basis the 
effective mass equation for electrons and holes in heterosystems for 
arbitrary interface orientation is obtained. Using this basis one can 
exclude off-diagonal matrix elements of the interface scattering potential 
and in the explicit form introduce into the theory position-dependent 
parameters, that describe bulk material (electron energy at 
${\bf k} = 0$ and band parameters) and can be taken from experiment. 
This approach is close to the theories developed by Karavaev and 
Tikhodeev \cite{3} and Foreman \cite{8}, \cite{10}.

To obtain the effective $\bf kp$-Hamiltonians with position-dependent 
parameters in $A_3B_5$-heterointerface systems a modified method 
of invariants is suggested. Since the presence of interface disturbs 
translational symmetry of the crystal and leads 
to scattering of the particles, 
changing their wave vector from $\bf k$ to ${\bf k}^\prime$, the effective 
Hamiltonian of the heterosystem, in contrast to the bulk material, should 
be considered dependent on these two variables. 
Besides that the effective $\bf kp$-Hamiltonian has 
to depend on the interface orientation, and therefore it should be 
described by one more independent variable $\bf n$ (the normal 
to the interface). Using this approach we obtained for arbitrary interface 
orientation the effective $\bf kp$-Hamiltonians with position-dependent 
band parameters for the $\Gamma_1$ and $\Gamma_6$ 
conduction bands, $\Gamma_{15}$ and $\Gamma_8$ valence bands, 
and also $\Gamma_8 {\oplus} \Gamma_7$ two-band effective 
$\bf kp$-Hamiltonian.

In the framework of this approach the effective $\bf kp$-Hamiltonian 
with position-dependent parameters can be represented as a sum 
of two parts: bulk-like part $H^{0(\alpha)}$ and short range part 
$H^{S(\alpha)}$. 
The bulk-like part, along with the usual terms, that are present 
in the Hamiltonian of a homogenous material, also includes additional 
quadratic-in-$\bf k$ terms. They arise due to non-commutativity of 
the momentum operator and band parameters and vanish in the bulk 
material. Analysis shows that these terms describe effective spin-orbit 
coupling at the interface and lead to spin-dependent effects of electron 
and hole interface scattering.

Besides that the bulk-like part of the $\bf kp$-Hamiltonian 
$H^{0(\alpha)}$ can include non-relativistic linear-in-$\bf k$ term, 
that appears due to variation of the Bloch functions across the interface. 
The short-range part  $H^{S(\alpha)}$ 
includes terms proportional to $n$, $n^2$, $nk$, $nk^2$, 
$\ldots$. The terms proportional to the first order of $\bf n$, 
which characterize the uniform electric field localized in the interface region, 
are the most important. In particular, in the $\Gamma_8$ 
band these corrections lead to a well known effect of heavy-light 
holes mixing at the interface. Analysis shows that $kn$ and $kn^2$ type 
terms are responsible for linear-in-$\bf k$ Rashba-like terms in the electron and hole
subbands dispersion in quantum wells. Influence of these terms on the 
effect can exceed the contribution of the cubic-in-$\bf k$ terms, 
which exist in the bulk Hamiltonian of the 
$A_3B_5$-semiconductors due to absence of the inversion symmetry.

The effective Hamiltonians derived in this paper allow one to 
obtain the correct boundary conditions for electron and hole 
envelope functions for arbitrary interface orientation. The additional 
terms, which were introduced into the Hamiltonians in this paper, 
modify usually used boundary conditions. However, some short range 
corrections (for example $n^2$, $nk$, $nk^2$ type terms) are small and
can be easily taken into account by the perturbation theory in the 
framework of the envelope function approximation.

\begin {center}
\bf
ACKNOWLEDGMENT
\end {center}

We would like to thank Prof. E.L. Ivchenko for some helpful discussions.

\begin {center}
\bf
APPENDIX A: HIGH ORDERS SHORT RANGE PARTS OF THE HAMILTONIANS
\end {center}

\begin {center}
\bf
A1. $\Gamma_1$ band Hamiltonian
\end {center}

$$
H^{S(\Gamma_1)}_{n^2} ( {\bf R})=
\frac{\hbar^2}{2m_0 a_0} g ({\bf n} \cdot {\bf n})
\delta ({\bf nR}),
$$

$$
H^{S(\Gamma_1)}_{kn} (\hat{\bf k} , {\bf R})=
i\frac{\hbar^2}{2m_0}l
\left( \left[ \hat{\bf k} \delta \right] \cdot {\bf n} \right) ,
$$

$$
H^{S(\Gamma_1)}_{kn^2} (\hat{\bf k} , {\bf R})=
i\frac{\hbar^2}{2m_0}q
\left( \left[ \hat{\bf k} \delta \right] \cdot {\bf N} \right) .
$$

\par\noindent
Here $\left[ \hat{\bf k} \delta \right] =\hat{\bf k} \delta({\bf nR}) 
-\delta({\bf nR})\hat{\bf k}$.

\begin {center}
\bf
A2. $\Gamma_6$ band Hamiltonian
\end {center}

$$
H^{S(\Gamma_6)}_{n^2} ( {\bf R})=
\frac{\hbar^2}{2m_0 a_0} \eta ({\bf n} \cdot {\bf n})
I \delta ({\bf nR}) ,
$$

$$
H^{S(\Gamma_6)}_{kn} (\hat{\bf k} , {\bf R})=
\frac{\hbar^2}{2m_0}
\Biggl[ i\lambda_1 \left( \left[ \hat{\bf k} \delta \right] 
\cdot {\bf n} \right) I +
\lambda_2 \left( \left[ \left\{  \hat{\bf k} \delta \right\}
\times {\bf n} \right] \cdot {\bf \sigma}\right) \Biggr],
$$

$$
\begin {array}{c}
H^{S(\Gamma_6)}_{kn^2} (\hat{\bf k} , {\bf R})=
\frac{\hbar^2}{2m_0}
\Biggl[ i\mu_1 \left( \left[ \hat{\bf k} \delta \right] 
\cdot {\bf N} \right) I+
\mu_2 \sum\limits_i \left\{ {\hat  k}_i \delta \right\} 
(n^2_{i+1}- n^2_{i+2}) \sigma_i +\\
\mu_3 \left( \left[ \left\{ \hat{\bf k} \delta \right\}
\times {\bf N} \right] \cdot {\bf \sigma}\right)
\Biggr].
\end {array}
$$

\par\noindent
Here $\left\{ \hat{\bf k} \delta \right\} = \frac{1}{2}
\left[ \hat{\bf k} \delta({\bf nR}) + 
\delta({\bf nR})\hat{\bf k}\right]$, $I$ is a unit matrix $2 \times 2$, 
$\sigma_i$ are Pauli matrices. If we neglect spin-orbit mixing
$\eta = g$, $\lambda_1 =l$, $\mu_1 = q$ and 
$\lambda_2 = \mu_2 = \mu_3 = 0$.

\begin {center}
\bf
A3. $\Gamma_{15}$ band Hamiltonian
\end {center}

$$
\begin {array} {c}
H^{S(\Gamma_{15})}_{n^2} ( {\bf R})=
\frac{\hbar^2}{2m_0 a_0}
\Biggl[ s_1 ({\bf n} \cdot {\bf n}) I
\delta ({\bf nR}) + 6s_2 
\sum\limits_i (n^2_i - \frac{1}{3}) J^2_i \delta ({\bf nR})+\\
12 s_3 \sum\limits_i N_i \left\{J_{i+1}J_{i+2}\right\}
\delta ({\bf nR}) 
\Biggr],
\end{array}
$$

$$
\begin {array}{c}
H^{S(\Gamma_{15})}_{kn} (\hat{\bf k} , {\bf R})=
\frac{\hbar^2}{2m_0}
\Biggl[ it_1 \left( \left[ \hat{\bf k} \delta \right] 
\cdot {\bf n} \right) I +
i6t_2 \sum\limits_i 
\left( \left[  {\hat k}_i \delta \right] n_i -
\frac{1}{3} \left( \left[{\hat {\bf k}} \delta \right] \cdot {\bf n}
\right) \right) J^2_i +\\
i 12 t_3 \sum\limits_i \left\{ \left[{\hat k}_i \delta \right] 
n_{i+1} \right\}
\left\{J_i J_{i+1}\right\}
+3t_4 \left( \left[ \left\{ \hat{\bf k} \delta \right\}
\times {\bf n} \right] \cdot {\bf J} \right)
\Biggr],
\end {array}
$$

$$
\begin {array}{c}
H^{S(\Gamma_{15})}_{kn^2} (\hat{\bf k} , {\bf R})=
\frac{\hbar^2}{2m_0}
\Biggl[ i 2 \sqrt 3 w_1 
\sum\limits_i \left[ {\hat k}_i \delta \right] ({\bf n} \cdot {\bf n})
\left\{J_{i+1}J_{i+2} \right\}+\\
i 2 \sqrt 3  w_2 \sum\limits_i \left[ {\hat k}_i \delta \right] 
\left( 2n^2_i- n^2_{i+1} - n^2_{i+2} \right) 
\left\{ J_{i+1}J_{i+2} \right\}+\\
3 w_3 \sum\limits_i \left\{ {\hat k}_i \delta \right\}
\left( n^2_{i+1} - n^2_{i+2} \right) J_i+
iw_4 \left( \left[ \hat{\bf k} \delta \right] 
\cdot {\bf N} \right) I +\\
i6w_5 \sum\limits_i 
\left( \left[  {\hat k}_i \delta \right] N_i -
\frac{1}{3} \left( \left[{\hat {\bf k}} \delta \right] \cdot {\bf N}
\right) \right) J^2_i +\\
i 12 w_6 \sum\limits_i \left\{ \left[{\hat k}_i \delta \right] 
N_{i+1}\right\}
\left\{J_i J_{i+1}\right\}
+3 w_7 \left( \left[ \left\{ \hat{\bf k} \delta \right\}
\times {\bf N} \right] \cdot {\bf J}\right)
\Biggr].
\end {array}
$$

\par\noindent
Here $I$ is a unit matrix $3 \times 3$, $J_i$ are matrices of the 
angular momentum $J = 1$.

\begin {center}
\bf
A4. $\Gamma_8$ band Hamiltonian
\end {center}

$$
\begin {array} {c}
H^{S(\Gamma_8)}_{n^2} ( {\bf R})=
\frac{\hbar^2}{2m_0 a_0}
\Biggl[ \xi_1 ({\bf n} \cdot {\bf n})
I \delta ({\bf nR}) + 2 \xi_2 
\sum\limits_i (n^2_i - \frac{1}{3}) J^2_i \delta ({\bf nR})+\\
4 \xi_3 \sum\limits_i N_i \left\{J_{i+1}J_{i+2}\right\}
\delta ({\bf nR}) 
\Biggr],
\end{array}
$$

$$
\begin {array}{c}
H^{S(\Gamma_8)}_{kn} (\hat{\bf k} , {\bf R})=
\frac{\hbar^2}{2m_0}
\Biggl[ i\tau_1 \left( \left[ \hat{\bf k} \delta \right] 
\cdot {\bf n} \right) I +
i 2 \tau_2 \sum\limits_i 
\left( \left[  {\hat k}_i \delta \right] n_i -
\frac{1}{3} \left( \left[{\hat {\bf k}} \delta \right] \cdot {\bf n}
\right) \right) J^2_i +\\
i 4 \tau_3 \sum\limits_i \left\{ \left[{\hat k}_i \delta \right] 
n_{i+1}\right\}
\left\{J_i J_{i+1}\right\}
+2 \tau_4 \left( \left[ \left\{ \hat{\bf k} \delta \right\}
\times {\bf n} \right] \cdot {\bf J}\right)+\\
8 \tau_5 \sum\limits_i {\left[ \left\{ \hat{\bf k} \delta \right\}
\times {\bf n} \right]}_i J^3_i +
\frac{4}{\sqrt 3}\tau_6  \sum\limits_i  
\left\{ \left\{ {\hat k}_{i+1} \delta\right\}
n_{i+2}\right\}
\left\{ J_i \left(J^2_{i+1}-J^2_{i+2}\right) \right\}
\Biggr],
\end {array}
$$

$$
\begin {array}{c}
H^{S(\Gamma_8)}_{kn^2} (\hat{\bf k} , {\bf R})=
\frac{\hbar^2}{2m_0}
\Biggl[
i \frac{2}{\sqrt 3}\chi_1 
\sum\limits_i \left[ {\hat k}_i \delta \right] 
({\bf n} \cdot {\bf n})
\left\{J_{i+1}J_{i+2} \right\}+\\
i \frac{2}{\sqrt 3} \chi_2 \sum\limits_i \left[ {\hat k}_i \delta \right]
\left( 2n^2_i- n^2_{i+1} - n^2_{i+2} \right)
\left\{J_{i+1}J_{i+2} \right\}+\\
2 \chi_3 \sum\limits_i \left\{ {\hat k}_i \delta \right\}
\left( n^2_{i+1} - n^2_{i+2} \right) J_i+
i\chi_4 \left( \left[ \hat{\bf k} \delta \right] 
\cdot {\bf N} \right) I +\\
i 2 \chi_5 \sum\limits_i 
\left( \left[  {\hat k}_i \delta \right] N_i -
\frac{1}{3} \left( \left[{\hat {\bf k}} \delta \right] \cdot {\bf N}
\right) \right) J^2_i +
i 4 \chi_6 \sum\limits_i \left\{ \left[{\hat k}_i \delta \right] 
N_{i+1}\right\}
\left\{J_i J_{i+1}\right\}+\\
2 \chi_7 \left( \left[ \left\{ \hat{\bf k} \delta \right\}
\times {\bf N} \right] \cdot {\bf J}\right)+
\frac{4}{\sqrt 3}\chi_8 \sum\limits_i \left\{ {\hat k}_i \delta \right\} 
\left( {\bf n}\cdot {\bf n} \right)
\left\{J_i \left(J^2_{i+1}-J^2_{i+2}\right)
\right\}+\\
\frac{4}{\sqrt 3}\chi_9 \sum\limits_i \left\{ {\hat k}_i \delta \right\}
\left(2n^2_i-n^2_{i+1}-n^2_{i+2} \right)
\left\{J_i \left(J^2_{i+1}-J^2_{i+2}\right)
\right\} +
8 \chi_{10} \sum\limits_i \left\{ {\hat k}_i \delta \right\}
\left( n^2_{i+1} - n^2_{i+2} \right) J^3_i+\\
\frac{4}{\sqrt 3}\chi_{11} \sum\limits_i  \left\{ \left\{{\hat k}_{i+1} 
\delta \right\} N_{i+2}\right\}
\left\{J_i \left(J^2_{i+1}-J^2_{i+2}\right)
\right\} +
8 \chi_{12}  \sum\limits_i {\left[ \left\{ \hat{\bf k} \delta \right\}
\times {\bf N} \right]}_i J^3_i \Biggr].
\end {array}
$$

\par\noindent
Here $I$ is a unit matrix $4 \times 4$, $J_i$ are matrices of the 
angular momentum $J = 3/2$. If we neglect spin-orbit mixing
$\xi_i = s_i$ ($i = 1, 2, 3$), $\tau_i = t_i$ ($i = 1, 2, 3, 4$),
$\chi_i = w_i$ ($i = 1, \ldots, 7$),
$\tau_5 = \tau_6 = \chi_8 = \cdots =\chi_{12} = 0$.

\begin {center}
\bf
A5. $\Gamma_8 {\oplus}\Gamma_7$ band Hamiltonian
\end {center}

$$
H^{S(\Gamma_8 {\oplus} \Gamma_7)}_{\xi} ({\hat {\bf k}}, {\bf R})=
\left[
\begin {array} {cc}
H^{S(\Gamma_8)} _{\xi}({\hat {\bf k}}, {\bf R})&
H^{S(\Gamma_8 \Gamma_7)} _{\xi}({\hat {\bf k}}, {\bf R})\\
 H^{S(\Gamma_8 \Gamma_7)^+} _{\xi}({\hat {\bf k}}, {\bf R})& 
H^{S(\Gamma_7)} _{\xi}({\hat {\bf k}}, {\bf R})
\end {array}
\right],
$$

\par\noindent
Here $\xi = n^2, kn, kn^2$; 
$H^{S(\Gamma_8)}_{\xi} (\hat{\bf k} ,{\bf R})$ coincide 
with the corresponding matrices of the $\Gamma_8$ band 
(see Appendix A4), in which
$\tau_5 = \tau_6 = \chi_8 = \cdots =\chi_{12} = 0$.

$$
H^{S(\Gamma_7)}_{n^2} ( {\bf R}) = 
\frac{\hbar^2}{2m_0 a_0}
\xi_1 ({\bf n} \cdot {\bf n})
I \delta ({\bf nR}),
$$

$$
H^{S(\Gamma_7)}_{kn} (\hat{\bf k} , {\bf R})=
\frac{\hbar^2}{2m_0}
\Biggl[ i\tau_1 \left( \left[ \hat{\bf k} \delta \right] 
\cdot {\bf n} \right) I +
2 \tau_4 \left( \left[ \left\{ \hat{\bf k} \delta \right\}
\times {\bf n} \right] \cdot {\bf \sigma}\right)
\Biggr],
$$

$$
\begin {array}{c}
H^{S(\Gamma_7)}_{kn^2} (\hat{\bf k} , {\bf R})=
\frac{\hbar^2}{2m_0}
\Biggl[
2 \chi_3 \sum\limits_i \left\{ {\hat k}_i \delta \right\}
\left( n^2_{i+1} - n^2_{i+2} \right) \sigma_i+\\
i\chi_4 \left( \left[ \hat{\bf k} \delta \right] 
\cdot {\bf N} \right) I +
2 \chi_7 \left( \left[ \left\{ \hat{\bf k} \delta \right\}
\times {\bf N} \right] \cdot {\bf \sigma}\right)
 \Biggr],
\end {array}
$$

$$
\begin {array}{c}
H^{S(\Gamma_8 \Gamma_7)}_{n^2} ( {\bf R}) = 
\frac{\hbar^2}{2m_0 a_0}
\Biggl[ 
\sqrt 2 \xi_2 \left[ (2n^2_z-n^2_x-n^2_y) I^{\Gamma_{12}}_1
+ \sqrt 3 (n^2_x-n^2_y) I^{\Gamma_{12}}_2 \right] \delta ({\bf nR}) +\\
\sqrt 6 \xi_3 \sum\limits_i N_i I^{\Gamma_{15}}_i
\delta ({\bf nR}) 
\Biggr],
\end {array}
$$

$$
\begin {array} {c}
H^{S(\Gamma_8 \Gamma_7)}_{kn} (\hat{\bf k} , {\bf R})=\\
\frac{\hbar^2}{2m_0}
\Biggl[ 
i \sqrt 2 \tau_2 \left[
\left(
2\left[ {\hat k}_z \delta \right]n_z-
\left[{\hat k}_x \delta \right]n_x-
\left[ {\hat k}_y \delta \right] n_y 
\right) 
I^{\Gamma_{12}}_1+
 \sqrt 3 
\left( 
\left[{\hat k}_x \delta \right]n_x-
\left[{\hat k}_y \delta \right]n_y 
\right) 
I^{\Gamma_{12}}_2 
\right]+\\
i \sqrt 6 \tau_3 \sum\limits_i 
\left\{ \left[ {\hat k}_{i+1} \delta \right] n_{i+2} \right\}
 I^{\Gamma_{15}}_{i}+
\frac{1}{\sqrt 2} \tau_4 \sum\limits_i 
{\left[
\left\{{\hat{ \bf k}} \delta \right\} \times {\bf n}
\right]}_i
 I^{\Gamma_{25}}_{i}
\Biggr],
\end {array}
$$

$$
\begin {array} {c}
H^{S(\Gamma_8 \Gamma_7)}_{kn^2} (\hat{\bf k} , {\bf R})=
\frac{\hbar^2}{2m_0}
\Biggl[ 
i \frac{1}{\sqrt 2}\chi_1 \sum\limits_i 
\left[{\hat k}_i \delta \right] \left({\bf n} \cdot {\bf n} \right)
 I^{\Gamma_{15}}_{i}+\\
i \frac{1}{\sqrt 2} \chi_2 \sum\limits_i 
\left[{\hat k}_i \delta \right]  (2n^2_i-n^2_{i+1}-n^2_{i+2})
 I^{\Gamma_{15}}_{i}
+\frac{1}{\sqrt 2}\chi_3 \sum\limits_i 
\left\{{\hat k}_i \delta \right\}  (n^2_{i+1}-n^2_{i+2})
 I^{\Gamma_{25}}_{i}+\\
i \sqrt 2 \chi_5 \left[
\left(
2\left[ {\hat k}_z \delta \right]N_z-
\left[{\hat k}_x \delta \right]N_x-
\left[ {\hat k}_y \delta \right] N_y 
\right) 
I^{\Gamma_{12}}_1+
 \sqrt 3 
\left( 
\left[{\hat k}_x \delta \right]N_x-
\left[{\hat k}_y \delta \right]N_y 
\right) 
I^{\Gamma_{12}}_2 
\right]+\\
i \sqrt 6 \chi_6 \sum\limits_i 
\left\{ \left[ {\hat k}_{i+1} \delta \right] N_{i+2} \right\}
 I^{\Gamma_{15}}_{i}
+\frac{1}{\sqrt 2}\chi_7 \sum\limits_i 
{\left[
\left\{{\hat{ \bf k}} \delta \right\} \times {\bf N}
\right]}_i
 I^{\Gamma_{25}}_{i}
\Biggr].
\end {array}
$$

\begin {center}
\bf
APPENDIX B: MATRICES $I^{\Gamma_\alpha}_i$
\end {center}

Matrices  $I^{\Gamma_\alpha}_i$, used to obtain the 
$ H^{(\Gamma_8\Gamma_7)} ({\hat {\bf k}}, {\bf R})$ 
block of the $\Gamma_8 {\oplus} \Gamma_7$ two-band 
Hamiltonian are:

$$
I^{\Gamma_ {12}}_1=
\left[
\begin {array} {cc}
0&0\\
-1&0\\
0&1\\
0&0
\end {array}
\right];
$$

$$
I^{\Gamma_ {12}}_2=
\left[
\begin {array} {cc}
0&1\\
0&0\\
0&0\\
- 1&0
\end {array}
\right];
$$

$$
I^{\Gamma_ {15}}_x=
\left[
\begin {array} {cc}
- i &0\\
0&i\sqrt 3\\
-i\sqrt 3&0\\
0&i 
\end {array}
\right];
$$

$$
I^{\Gamma_ {15}}_y=
\left[
\begin {array} {cc}
1&0\\
0&- \sqrt 3\\
- \sqrt 3&0\\
0&1
\end {array}
\right];
$$

$$
I^{\Gamma_ {15}}_z=
\left[
\begin {array} {cc}
0&-i2\\
0&0\\
0&0\\
-i2 &0
\end {array}
\right];
$$

$$
I^{\Gamma_ {25}}_x=
\left[
\begin {array} {cc}
\sqrt {3}&0\\
0&1\\
-1&0\\
0&-\sqrt {3}
\end {array}
\right];
$$

$$
I^{\Gamma_ {25}}_y=
\left[
\begin {array} {cc}
-i \sqrt {3}&0\\
0&-i \\
-i\\
0&-i\sqrt {3}
\end {array}
\right];
$$

$$
I^{\Gamma_ {25}}_z=
\left[
\begin {array} {cc}
0&0\\
- 2&0\\
0&- 2\\
0&0
\end {array}
\right];
$$


\begin {thebibliography} {99}

\bibitem[*]{*}Electronic mail: SaulTailor@mail.nevalink.ru

\bibitem{1}L. Leibler, Phys. Rev. B {\bf 12}, 4443 (1975).

\bibitem{2}G. Bastard, {\it Wave Mechanics Applied to 
Semiconductor Heterostructures}
 (Wiley, New York, 1988).

\bibitem{3}G.F. Karavaev and Yu.S. Tikhodeev, Fiz. Tekh. Poluprovodn. 
{\bf 25}, 1237 (1991) [Sov. Phys. Semicond. {\bf 25}, 745 (1991)]. 

\bibitem{4}M.G. Burt, J. Phys. Condens. Matter {\bf 4}, 6651 (1992).

\bibitem{5}M.G. Burt, Phys. Rev. B {\bf 50}, 7518 (1994).

\bibitem{6}B.A. Foreman, Phys. Rev. B {\bf 52}, 12241 (1995).

\bibitem{7} E.L. Ivchenko and G.E. Pikus, {\it Superlattices and Other 
Heterostructures. Symmetry and Optical Phenomena}, Springer 
Series in Solid-State Sciences Vol. 110 (Springer-Verlag, 1997).

\bibitem{8}B.A. Foreman, Phys. Rev. B {\bf 54}, 1909 (1996).

\bibitem{9}C.Y.-P. Chao and S.L. Chuang, Phys. Rev. B {\bf 43}, 7027 (1991). 

\bibitem{10}B.A. Foreman, Phys. Rev. B {\bf 48}, 4964 (1993).

\bibitem{11}E.E. Takhtamirov and V.A. Volkov, Semicond. Sci. 
Technol. {\bf 12}, 77 (1997).

\bibitem{12}F.S.A. Cavalcante, R.N. Costa Filho, J. Ribeiro Filho, 
C.A.S. de Almeida, and V.N. Freire, Phys. Rev. B {\bf 55}, 1326 (1997).

\bibitem{13}G.F. Glinskii and K.O. Kravchenko, Fiz. Tverd. 
Tela {\bf 40}, 85 (1998).

\bibitem{14}E.L. Ivchenko, A.Yu. Kaminski, and U. R{\"o}ssler, 
Phys. Rev. B {\bf 54}, 5852 (1996).

\bibitem{15}Y. Fu, M. Willander, E.L. Ivchenko, and A.A. Kiselev, 
Phys. Rev. B {\bf 47}, 13498 (1993).

\bibitem{16}E.L. Ivchenko, A.A. Kiselev, Y. Fu, M. Willander,
Phys. Rev. B {\bf 50}, 7747 (1994).

\bibitem{17}O. Krebs, W. Seidel, J.P. Andr{\'e}, D. Bertho, C. Jouanin, 
and P. Voisin, Semicond. Sci. Technol. {\bf 12}, 938 (1997).

\end {thebibliography}
\end{document}